\pdfoutput=1

\documentclass[11pt,twoside,a4paper,cmspaper,final,collab]{cms-tdr}
\def\svnVersion{268244}\def\cmsCernNo{2013-037}\def\cmsCernDate{\today}\def\cmsMessage{Published in the European Physical Journal C as \href{http://dx.doi.org/10.1140/epjc/s10052-014-3129-3}{\doi{10.1140/epjc/s10052-014-3129-3.}}}

\begin{document}\cmsNoteHeader{SMP-13-001}

\hyphenation{had-ron-i-za-tion}
\hyphenation{cal-or-i-me-ter}
\hyphenation{de-vices}
\RCS$Revision: 261738 $
\RCS$HeadURL: svn+ssh://svn.cern.ch/reps/tdr2/papers/SMP-13-001/trunk/SMP-13-001.tex $
\RCS$Id: SMP-13-001.tex 261738 2014-09-24 19:17:04Z alverson $
\newlength\cmsFigWidth
\ifthenelse{\boolean{cms@external}}{\setlength\cmsFigWidth{0.85\columnwidth}}{\setlength\cmsFigWidth{0.4\textwidth}}
\ifthenelse{\boolean{cms@external}}{\providecommand{\cmsLeft}{top}}{\providecommand{\cmsLeft}{left}}
\ifthenelse{\boolean{cms@external}}{\providecommand{\cmsRight}{bottom}}{\providecommand{\cmsRight}{right}}
\newcommand{\mgg}{\ensuremath{m_{\Pgg\Pgg}}\xspace}
\newcommand{\ptgg}{\ensuremath{\pt^{\Pgg\Pgg}}\xspace}
\newcommand{\dphigg}{\ensuremath{\Delta \phi_{\Pgg\Pgg}}\xspace}
\newcommand{\startheta}{\ensuremath{\theta^*}\xspace}
\newcommand{\RESBOS} {{\textsc{resbos}}\xspace}
\newcommand{\DIPHOX} {{\textsc{diphox}}\xspace}
\newcommand{\GENERATORGTOMC} {{\textsc{gamma2mc}}\xspace}
\newcommand{\TWOGNNLO} {{\textsc{2\Pgg nnlo}}\xspace}
\newcommand{\MYETONE}{\ensuremath{E_{\mathrm{T}}^{\Pgg\mathrm{1}}}\xspace}
\newcommand{\MYETTWO}{\ensuremath{E_{\mathrm{T}}^{\Pgg\mathrm{2}}}\xspace}
\ifthenelse{\boolean{cms@external}}{\providecommand{\cmsRow}[1]{\resizebox{\columnwidth}{!}{#1}}}{\providecommand{\cmsRow}[1]{\relax{#1}}}

\cmsNoteHeader{SMP-13-001} 
\title{Measurement of differential cross sections for the production of a pair of isolated photons in pp collisions at $\sqrt{s}=7\TeV$}

\titlerunning{Differential cross sections for isolated photon pairs}
\author{The CMS Collaboration}

\date{\today}

\abstract{
A measurement of differential cross sections for the production of a pair of isolated photons in proton-proton collisions at $\sqrt{s}=7\TeV$ is presented. The data sample corresponds to an integrated luminosity of 5.0\fbinv collected with the CMS detector. A data-driven isolation template method is used to extract the prompt diphoton yield.  The measured cross section for two isolated photons, with transverse energy above 40 and 25\GeV respectively, in the pseudorapidity range $\abs{\eta}<2.5$, $\abs{\eta}\notin[1.44,1.57]$ and with an angular separation $\Delta R > 0.45$, is $17.2 \pm 0.2\stat \pm 1.9\syst\pm 0.4\lum$\unit{pb}. Differential cross sections are measured as a function of the diphoton invariant mass, the diphoton transverse momentum, the azimuthal angle difference between the two photons, and the cosine of the polar angle in the Collins--Soper reference frame of the diphoton system. The results are compared to theoretical predictions at leading, next-to-leading, and next-to-next-to-leading order in quantum chromodynamics.
}

\hypersetup{%
pdfauthor={CMS Collaboration},%
pdftitle={Measurement of differential cross sections for the production of a pair of isolated photons in pp collisions at sqrt(s) = 7 TeV},%
pdfsubject={CMS},%
pdfkeywords={CMS, physics, diphoton, photon, measurement}}

\maketitle 

\section{Introduction}
\label{sec:Introduction}

The measurement of differential diphoton production cross sections offers an important test of both perturbative and non-perturbative quantum chromodynamics (QCD). At leading order (LO), diphotons are produced via quark-antiquark annihilation \cPq\cPaq\ $\to$ \cPgg\cPgg. At next-to-leading order (NLO), diphoton production also includes the quark-gluon channel, while next-to-next-to-leading order (NNLO) adds the gluon-gluon channel, which includes a box diagram and represents a non-negligible fraction of the total cross section. Diphoton production is sensitive to the emission of soft gluons in the initial state and to the non-perturbative fragmentation of quarks and gluons to photons in the final state. Due to this rich phenomenology, theoretical predictions are challenging especially in restricted regions of phase space.

Diphoton production constitutes the major source of background in the diphoton decay channel of the newly discovered Higgs boson \cite{DiscoveryATLAS, DiscoveryCMS, CMSLongPaper}, as well as to searches for physics beyond the standard model. New physics processes may also appear as non-resonant deviations from the predicted diphoton spectrum in events with large missing transverse energy, as in gauge-mediated SUSY breaking \cite{GMSB} or in models of universal extra dimensions \cite{UED}. Alternatively, some models predict narrow resonances, such as the graviton in the Randall--Sundrum model for warped extra dimensions \cite{RS,RS2}.

The most recent diphoton measurements were published by the CDF and D0 Collaborations \cite{diphotonCDF,diphotonD0} at the Tevatron and by the ATLAS Collaboration \cite{diphotonATLAS} at the LHC. This paper presents an update of a previous CMS measurement at $\sqrt{s}=7\TeV$ \cite{DiphotonPaper2010} and is based on the full 2011 data sample of 5.0\fbinv. It probes a phase space defined by a highly asymmetric selection for the transverse energy (\ET) of the two photons. The leading-order kinematic configuration where photons are produced back-to-back in the transverse plane is suppressed, enhancing the sensitivity to higher-order diagrams. The ratio of the NNLO to the LO prediction is increased by 20\% with respect to the previous CMS measurement \cite{DiphotonPaper2010}.

The main experimental challenge for the measurement of the diphoton cross section is distinguishing the ``prompt'' photon signal produced either directly or as a result of fragmentation from the background that arises mainly from energetic neutral mesons, predominantly \Pgpz\ and \Pgh\ mesons, inside jets. These mesons typically decay to two collimated photons that are reconstructed as a single photon candidate, which is referred to as ``non-prompt'' in this paper. The main features used to discriminate a prompt photon from a non-prompt one are the shape of the shower measured by the electromagnetic calorimeter (ECAL) \cite{QCD-10-019} and the isolation energy in a cone around the photon direction \cite{DiphotonPaper2010,CMS-QCD-10-037}. This information can be used to trigger on diphoton candidate events and, at the analysis level, to statistically evaluate the fraction of prompt diphoton candidates.

The particle flow (PF) event reconstruction \cite{CMS-PAS-PFT-10-002} consists in reconstructing and identifying each particle with an optimal combination of all sub-detector information. In this process, the identification of the particle type (photon, electron, muon, charged hadron, neutral hadron) plays an important role in the determination of the particle direction and energy.

In this analysis, the photon component of the PF isolation is used as the discriminating variable. The distributions of signal and background components are built from data and used in a maximum likelihood fit to estimate the signal fraction. An increased separation power with respect to previous results \cite{DiphotonPaper2010} is achieved by improving the identification and subtraction of the photon energy deposit in the isolation cone.

After a brief description of the CMS detector in Section~\ref{sec:CMS}, data and simulated samples are presented in Section~\ref{sec:Dataset}, and the photon reconstruction in Section~\ref{sec:Photons}. The diphoton signal is estimated as described in Section~\ref{sec:Purity}. The number of signal events is then corrected for inefficiencies and unfolded as described in Section~\ref{sec:Efficiencies}. Systematic uncertainties are assessed in Section~\ref{sec:Systematics}, and the differential cross sections are presented and compared to theoretical predictions in Section~\ref{sec:Theory}.

\section{The CMS detector}
\label{sec:CMS}
A detailed description of the CMS detector can be found elsewhere~\cite{bib-detector}.
Its central feature is a superconducting solenoid,
13\unit{m} in length and 6\unit{m} in diameter,
which provides an axial magnetic field of 3.8\unit{T}.
The bore of the solenoid is instrumented with both the tracker (TRK) and the calorimeters.
The steel flux-return yoke outside the solenoid is instrumented with gas-ionisation detectors
used to reconstruct and identify muons.
Charged-particle trajectories are measured by the silicon pixel and
strip tracker,
with full azimuthal ($\phi$) coverage within $\abs{\eta} < 2.5$, where the pseudorapidity
$\eta$ is defined as $\eta = -\ln[\tan(\theta/2)]$, with $\theta$ being the
polar angle of the trajectory of the particle with respect to the
counterclockwise beam direction.
A lead tungstate crystal electromagnetic calorimeter (ECAL) and a
brass/scintillator
hadron calorimeter (HCAL) surround the tracking volume and cover the region
$\abs{\eta} < 3$.
The ECAL barrel (EB) extends to $\abs{\eta} < 1.479$ while the ECAL endcaps (EE) cover the region $1.479 < \abs{\eta} < 3.0$.
A lead/silicon-strip preshower detector (ES) is located in front of the
ECAL endcap in the region $1.653 < \abs{\eta} < 2.6$.
The preshower detector includes two planes of silicon sensors measuring the
$x$ and $y$ coordinates of the impinging particles.
In the $(\eta, \phi)$ plane, and for $\abs{\eta} < 1.48$, the HCAL cells map onto $5\times5$
ECAL crystal arrays to form calorimeter towers projecting radially
outwards from points slightly offset from the nominal interaction point.
In the endcap, the ECAL arrays matching the HCAL cells contain fewer crystals.
A steel/quartz-fibre Cherenkov forward calorimeter extends the
calorimetric coverage to $\abs{\eta} < 5.0$.

\section{Data sample}
\label{sec:Dataset}

The data sample consists of proton-proton (\Pp\Pp) collision events collected at the LHC with the CMS detector in the year 2011, at a centre-of-mass energy ($\sqrt{s}$) of 7\TeV and
corresponding to an integrated luminosity of 5.0\fbinv.

Events are triggered~\cite{bib-detector} by requiring the presence of two photons with asymmetric transverse energy thresholds.
The \ET thresholds at trigger level are 26\,(18) and 36\,(22)\GeV on the leading (sub-leading) photon, depending on the running period.
Each candidate is required to satisfy either loose calorimetric identification requirements, based on the shape of the electromagnetic shower, or loose isolation conditions.
The trigger efficiency is evaluated using a tag-and-probe technique on $\Z \to \Pep\Pem$ events \cite{WZpaper}, with electrons treated as photons. The trigger efficiency for photons selected in this analysis is measured to be between 98.8\% and 100\% depending on the pseudorapidity and the interaction with the material in front of the ECAL. The total trigger efficiency is found to be constant over the data taking period.

Several samples of simulated events are used in the analysis to model signal and background processes. Drell--Yan+jets and \cPgg\cPgg+jets signal events are generated with \MADGRAPH 1.4.8 \cite{Madgraph5}. The \cPg\cPg $\to$ \cPgg\cPgg\ box signal process, \cPgg+jet, and QCD dijet background processes are generated with \PYTHIA 6.4.24 \cite{Pythia6}. For all simulated samples the CTEQ6L1 \cite{CTEQ6} parton distribution functions (PDFs) are used. All generated events are then processed with \PYTHIA (Z2 tune) \cite{UEpaper} for hadronization, showering of partons and the underlying event; a detailed simulation of the CMS detector based on \GEANTfour~\cite{Agostinelli:2002hh} is performed, and the simulated events are finally reconstructed using the same algorithms as used for the data.

The simulation includes the effects of in-time pileup (overlapping \Pp\Pp\ interactions within a bunch crossing) and out-of-time pileup (overlapping \Pp\Pp\ interactions from interactions happening in earlier and later bunch crossings) with a distribution matching that observed in data.

\section{Photon reconstruction and selection}
\label{sec:Photons}

\subsection{Photon reconstruction}
{\tolerance=800
Photon candidates are reconstructed from the energy deposits in the ECAL
by grouping its channels into superclusters~\cite{EGM-11-001}.
About half of the photons convert into an \Pep\Pem\ pair in the material in front of the ECAL. Conversion-track pairs are
reconstructed from a combination of Gaussian-sum filter (GSF) electron tracks \cite{2005_gsf_paper} and ECAL-seeded tracks
fit to a common vertex and then matched to the photon candidate.
The superclustering algorithms achieve an almost complete collection of the energy
of such converted photons.
In the barrel region, superclusters are formed from five-crystal-wide strips
in $\eta$, centred on the locally most energetic crystal (seed), and
have a variable extension in the azimuthal direction ($\phi$).
In the endcaps, where the crystals are arranged according to an $x$-$y$ rather than an $\eta$-$\phi$ geometry,
matrices of $5\times5$ crystals (which may partially overlap) around the most energetic crystals are merged
if they lie within a narrow $\phi$ road.
The photon candidates are reconstructed within the ECAL fiducial region $\abs{\eta}<2.5$ but excluding the
barrel-endcap transition regions $1.44 < \abs{\eta} < 1.57$. This exclusion of the barrel-endcap transition regions ensures containment of the shower of the selected photon candidate in either the ECAL barrel or one of the ECAL endcaps.
The fiducial region requirement is applied to the supercluster position (defined as the log-weighted barycentre of the supercluster's active channels) in the ECAL.

The photon energy is computed starting from the raw crystal energies measured in the ECAL.
In the region covered by the preshower detector the energy recorded in that sub-detector is added.
The variation of the crystal transparency during the run is continuously monitored and corrected
using a factor based on the change in response to light from a laser and light-emitting-diode based monitoring system.
The single-channel response of the ECAL is equalised by exploiting the $\phi$ symmetry of the energy flow,
the mass constraint on the energy of the two photons in decays of \Pgpz\ and \Pgh\ mesons,
and the momentum constraint on the energy of isolated electrons from \PW\ and \Z decays.
A correction factor compensates for the imperfect containment of the shower in the cluster crystals.
The absolute energy scale and the residual long term drifts in the response are
further corrected using $\Z \to \Pep\Pem$ decays~\cite{EGM-11-001}.

Interaction vertices are reconstructed from charged tracks and the vertex of the diphoton event is taken as the one with the largest sum of squared transverse momenta ($\Sigma \pt^2$) of the associated tracks.
The photon four-momentum is recalculated with respect to this vertex.
\par}

\subsection{Photon selection}

The photon candidates are first required to pass a sequence of filters that aim to remove beam backgrounds or identified detector issues and to satisfy more stringent criteria than the trigger requirements.
The preselection is based on the shape of the electromagnetic shower in the ECAL and on the degree of isolation of the photon (i.e. the amount of energy deposited in the vicinity of the photon).
The variables used are:
\begin{itemize}
\item Photon supercluster raw energy $E^\text{raw}_\mathrm{SC}$: the sum of the calibrated crystal energies;
\item Preshower energy $E^\mathrm{ES}_\mathrm{SC}$: the sum of the energy deposits reconstructed in the preshower detector (ES) and associated with the supercluster;
\item \RNINE: the energy sum of $3 \times 3$ crystals centred on the most energetic crystal in the supercluster divided by the raw energy of the supercluster;
\item $H/E$: the ratio of the energy deposited in HCAL that is inside a cone of size $\Delta R=\sqrt{\smash[b]{(\Delta \eta)^2 + (\Delta \phi)^2}}=0.15$ centred on the photon direction, to the supercluster energy;
\item $\sigma_{\eta \eta}$: the shower transverse extension along $\eta$ that is defined as:

\begin{equation}
    \label{eq:pho_showcov}
    \sigma^2_{\eta \eta} = \frac
    { \sum \left( \eta_i - \bar\eta \right )^2  w_i
    } {\sum w_i},
\end{equation}

\noindent where the sum runs over all elements of the $5\times 5$ matrix around the most energetic crystal in the supercluster,
and $\eta_i=0.0174\ \hat\eta_i$ in EB, $\eta_i=0.0447\ \hat\eta_i$ in EE with $\hat\eta_i$ denoting the index of the
$i$th crystal along the $\eta$ direction. The individual weights $w_i$ are given by
$w_i = \max \left( 0, 4.7 + \ln(E_i / E_{5\times5}) \right )$, where $E_i$ is the
energy of the $i$th crystal and $\bar\eta=\sum\eta_iE_i/\sum E_i$
is the weighted average pseudorapidity;

\item $\text{Iso}_\text{ECAL}^{0.3}$ (ECAL isolation): the scalar sum of the \ET of the deposits in the electromagnetic calorimeter lying inside a cone of size $\Delta R= 0.3 $, centred on the direction of the supercluster but excluding an inner cone of size 3.5 crystals and an $\eta$-slice region of 2.5 crystals;
\item $\text{Iso}_\text{HCAL}^{0.3}$ (hadronic calorimeter isolation): the scalar sum of the \ET of the deposits in the hadron calorimeter that lie inside a hollow cone of outer radius of size $\Delta R = $ 0.3 and inner radius of size $\Delta R = $ 0.15 in the $\eta$-$\phi$ plane, centred on the direction of the supercluster;
\item $\text{Iso}_\text{TRK}^{0.3}$ (tracker isolation): the scalar sum of the \pt of the tracks that are consistent with originating from the primary vertex in the event, and lie inside a hollow cone of outer radius of size $\Delta R = $ 0.3 and inner radius of size $\Delta R =  0.04$ in the $\eta$-$\phi$ plane, centred around a line connecting the primary vertex with the supercluster but excluding an $\eta$-slice region ($\Delta \eta = 0.015$).
\end{itemize}

The isolation requirements are kept loose because the isolation is used as the discriminating variable in the signal extraction procedure.
The selection criteria are defined to be slightly tighter than the trigger selection.
The shower shape variables in the simulation are corrected to compensate for their imperfect modeling, mainly connected with (a)~the simulation of effective readout noise in ECAL channels, (b)~the effect of overlapping energy deposits from collisions in adjacent bunch crossings, and (c) the description of the material budget in the detector geometry.
The correction factors are extracted from a sample of photons in $\Z \to \Pgmp\Pgmm\Pgg$ events, and validated as a function of \ET and $\eta$ in a sample of electrons from \Z boson decays.
The list of preselection criteria is presented in Table \ref{table_preselection}.

\begin{table*}[ht!bp]
\centering
\topcaption{List of requirements that a candidate has to satisfy to pass the analysis preselection.}
\label{table_preselection}
\begin{tabular}{l|l} \hline
Variable & Requirement \\ \hline \hline
Photon raw + preshower energy  &  $E^\text{raw}_\mathrm{SC} + E^\mathrm{ES}_\mathrm{SC}> 20\GeV$ \\ \hline
$H/E$ & if ($\RNINE > 0.9$): $H/E < 0.082$ (EB), 0.075 (EE) \\
\  & if ($\RNINE < 0.9$): $H/E < 0.075$ \\ \hline
$\sigma_{\eta \eta}$ & $0.001 <\sigma_{\eta \eta}< 0.014$ (EB), 0.034 (EE) \\ \hline
ECAL isolation in a $\Delta R$=0.3 cone & $\mathrm{Iso}_\text{ECAL}^{0.3}< 4\GeV$ (only if $\RNINE < 0.9$)\\ \hline
HCAL isolation in a $\Delta R$=0.3 cone & $\mathrm{Iso}_\text{HCAL}^{0.3}< 4\GeV$ (only if $\RNINE < 0.9$)\\ \hline
TRK isolation in a $\Delta R$=0.3 cone & $\mathrm{Iso}_\text{TRK}^{0.3}< 4\GeV$ (only if $\RNINE < 0.9$)\\ \hline
\end{tabular}
\end{table*}

The preselected photons must satisfy additional requirements to be considered as photon candidates.
These consist of the absence of reconstructed electron track seeds in the pixel detector which match the candidate's direction,
and a tighter selection on the hadronic leakage of the shower and  the $\sigma_{\eta \eta}$ shower shape variable.
The list of additional selection criteria is shown in Table \ref{tab:photon_selection}.
\begin{table}
\centering
\topcaption{List of additional requirements applied in the photon candidate selection.}
\label{tab:photon_selection}
\cmsRow{
\begin{tabular}{ll} \hline
Variable & Requirement \\ \hline
Matched pixel measurements & False \\
$H/E$ & $H/E < 0.05$ \\
$\sigma_{\eta \eta}$ & $\sigma_{\eta \eta}< 0.011$ (EB), 0.030 (EE) \\ \hline
\end{tabular}
}
\end{table}

In the simulation, prompt photons are defined as candidates satisfying the analysis selection requirements and geometrically matched to an isolated generator-level photon, either directly produced or originating from a fragmentation process. The generator-level isolation is defined as the \pt sum of stable particles in a cone of size $\Delta R= 0.4 $, and is required to be less than 5\GeV.

\section{Signal yield determination}
\label{sec:Purity}

The diphoton signal is extracted from events containing two photon candidates with transverse energy greater than 40\,(25)\GeV for the leading~(sub-leading) photon, and with a separation of $\Delta R > 0.45$. If more than two photon candidates are selected, the two with highest \ET are retained. The minimum separation requirement ensures that the energy deposit of one photon does not enter the isolation cone centered on the other one.
The signal fraction is statistically separated from jets misidentified as photons by means of a binned maximum likelihood fit that uses the photon component of the PF isolation as the discriminating variable.

The diphoton signal is then studied as a function of the diphoton invariant mass \mgg, the diphoton transverse momentum \ptgg, the azimuthal angle difference \dphigg between the two photons, and the cosine of the polar angle \startheta in the Collins--Soper frame of the diphoton system \cite{CollinsSoper}. A maximum likelihood fit is performed for each bin of the distributions in the above variables.

\subsection{\label{sec:PFIsolation}Particle flow isolation}

The photon component of the PF isolation ($\text{Iso}$) is used to discriminate signal from background. The choice of the isolation variable is optimized to obtain the smallest total uncertainty of the measured cross section. This variable is computed, in a cone of size $\Delta R= 0.4 $ around each selected photon candidate, as the \ET sum of photons reconstructed with the PF algorithm \cite{CMS-PAS-PFT-10-002}. The PF isolation deals more effectively with cases of overlapping particles than the calorimetry-based isolation.

{\tolerance=800
When calculating the isolation, the energy deposited by the selected photon candidate is subtracted by removing from the cone the area where the photon is expected to have deposited its energy (``footprint''), since photon energy leaking into the cone could bias the isolation sum. This is done on an event-by-event basis relying on simple geometrical considerations. The directions of the momenta of reconstructed photon candidates around the selected photon are extrapolated from the interaction vertex to the inner surface of the ECAL, and whenever they overlap with a crystal belonging to the supercluster these photon candidates are removed from the isolation sum. For the matching between the propagated trajectory and the crystal front width, a tolerance of 25\% of the face size is applied.
\par}

This procedure does not use any generator-level information and can therefore be applied in both data and simulated events.

The pile-up introduces a spurious correlation between the two candidate photons' isolation sums. For this reason the PF isolation sums for both photons are corrected, event by event, for the presence of pile-up with a factor proportional to the average pile-up energy density ($\rho$) calculated with \textsc{FastJet} \cite{fastjet}.

\subsection{Template construction}

{\tolerance=800
The diphoton signal is extracted through a two-dimensional binned maximum likelihood fit that uses the isolation of the two selected photon candidates as discriminating variables. Different templates are built for the prompt-prompt ($f_{pp}$), prompt-non-prompt ($f_{pn}$), non-prompt-prompt ($f_{np}$), and non-prompt-non-prompt ($f_{nn}$) components in the ($\text{Iso}_1$, $\text{Iso}_2$) plane, where $\text{Iso}_1$ and $\text{Iso}_2$ represent the isolation variables for the two selected photon candidates in the event.
The probability distribution function has the following form:
\begin{equation}
\begin{aligned}
\mathcal{P}_{2D}(\text{Iso}_1,\text{Iso}_2) &=  f_{pp} T_{pp}(\text{Iso}_1,\text{Iso}_2) + f_{pn} T_{pn}(\text{Iso}_1,\text{Iso}_2) \\
\  & + f_{np} T_{np}(\text{Iso}_1,\text{Iso}_2) + f_{nn}T_{nn}(\text{Iso}_1,\text{Iso}_2)
\end{aligned}
\end{equation}
where $T_{kk}(\text{Iso}_1,\text{Iso}_2)$ is the function describing the isolation distribution (template) for the component $f_{kk}$.
Techniques have been developed to extract the templates from data to avoid possible biases coming from an imperfect modeling of the events in the simulation. Samples of events where at least one photon passes the photon selection are used to create prompt-prompt, prompt-non-prompt, non-prompt-prompt and non-prompt-non-prompt templates with high statistical precision, as described in the following.
\par}

The ``random cone'' technique is used to extract the prompt photon template with high statistical accuracy. In this procedure we compute the isolation energy in a region separated from the candidate photon. Starting from the photon $(\eta,\phi)$ axis, a new axis is defined at the same pseudorapidity $\eta$ but with a random separation in azimuthal angle $\phi_{RC}$ between 0.8 and $2\pi - 0.8$ radians from the photon $\phi$.
This new axis is used to define the random cone provided that no jet with $\pt > \text{20}$\GeV or photon or electron with $\pt > \text{10}$\GeV is reconstructed within $\Delta R < \text{0.8}$ and no muon is reconstructed within $\Delta R < \text{0.4}$ from this axis. In the case where the new axis does not meet these requirements, a new azimuthal angle is generated. The isolation energy, which is defined as the energy collected in a cone of size $\Delta R < \text{0.4}$ about the new axis once the fraction corresponding to the area of the photon supercluster has been removed, is then used to populate the prompt photon template.

The distribution of the template variable has been studied in \Z $\to$ \Pep\Pem\ events and found to be in agreement with the template built with the random cone technique.

The background (non-prompt) template cannot be defined by simply inverting the photon preselection, because the candidates entering the analysis, i.e. fulfilling the preselection requirements, have ``photon-like'' characteristics, while the set of candidates not fulfilling the photon preselection criteria includes a large number of genuine jets. To avoid this bias, the candidates selected to populate the non-prompt photon template are chosen from those that fulfil all the photon selection criteria, except the $\sigma_{\eta \eta}$ shower shape, which is not strongly correlated with the isolation variable as a result of the footprint removal technique described in the previous Section. The events in a ``sideband'' close to the photon selection criterium are used to populate the non-prompt photon template. The sideband is defined as $0.011 < \sigma_{\eta \eta} < 0.014$ for candidates reconstructed in the ECAL barrel and $0.030 < \sigma_{\eta \eta} < 0.034$ for candidates reconstructed in the ECAL endcaps.

\begin{figure}[htb]
\centering
\includegraphics[width=0.49\textwidth]{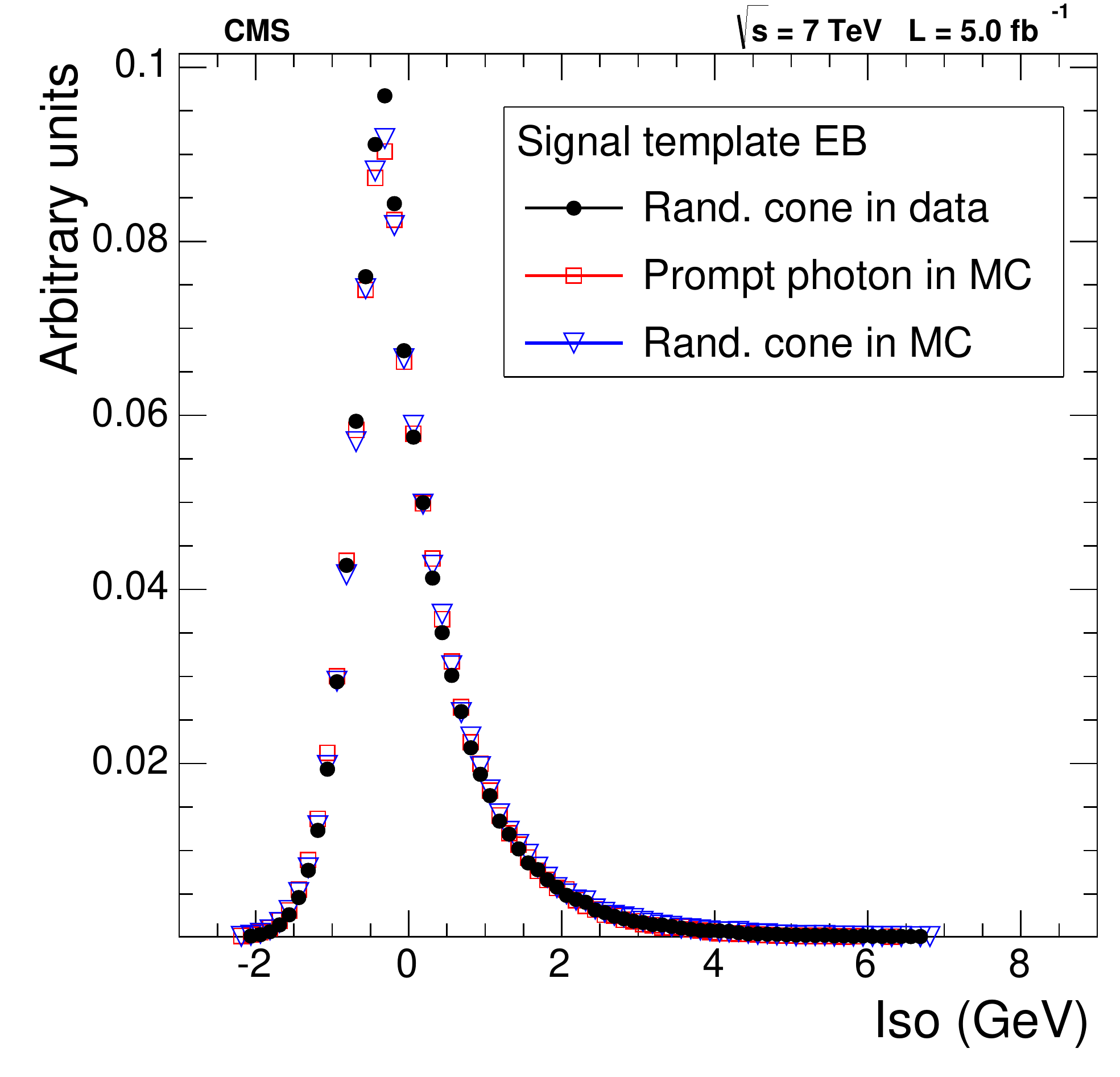} \hfill
\includegraphics[width=0.49\textwidth]{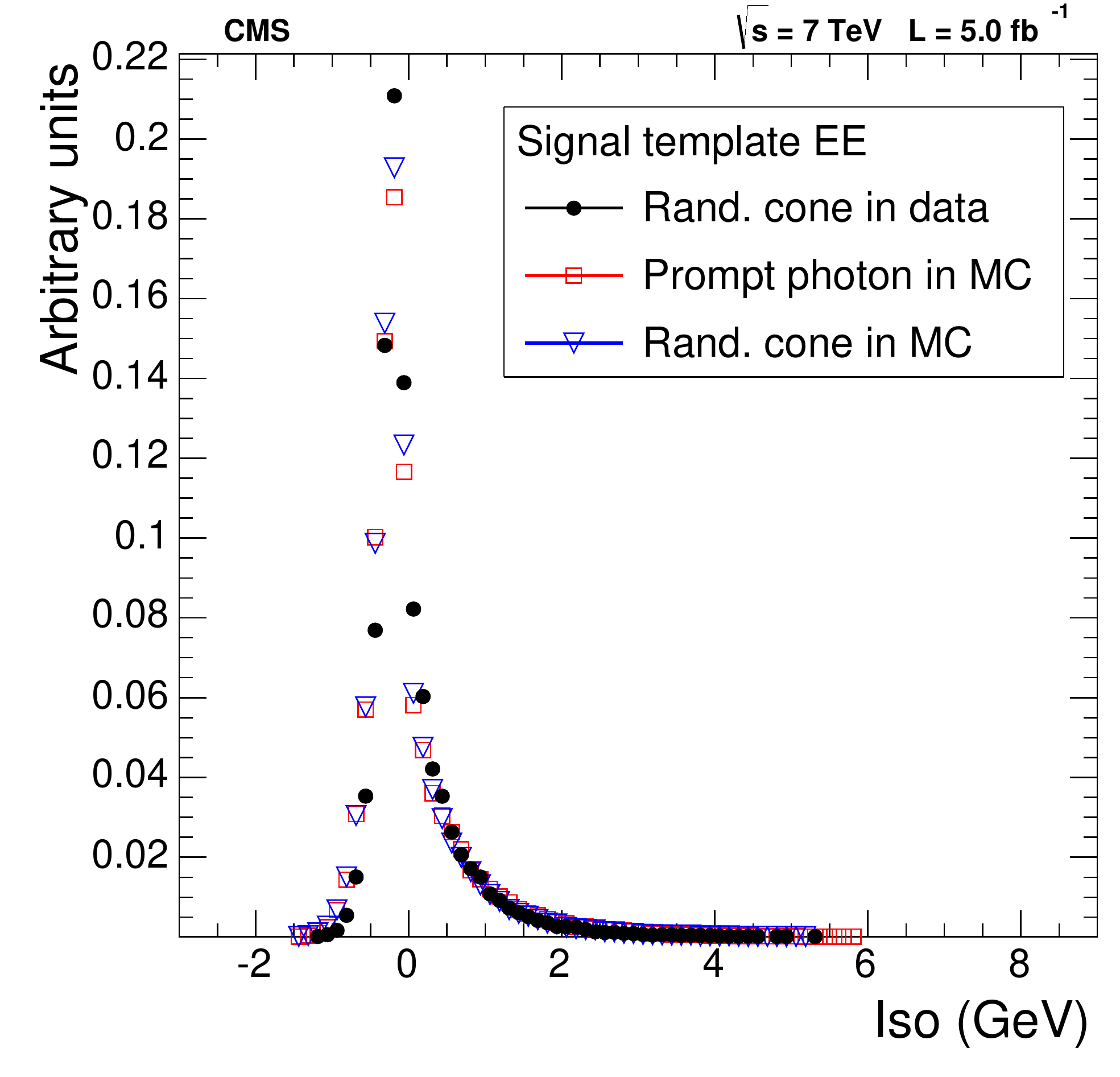}
\caption{Comparison of prompt photon templates in data and simulation: prompt photons in the simulation (squares), prompt photon templates extracted with the random cone technique from simulation (triangles) and from data (dots); (\cmsLeft) candidates in the ECAL barrel, (\cmsRight) candidates in the ECAL endcaps. All histograms are normalized to unit area.}
\label{fig:templateSignal}
\end{figure}

\begin{figure}[htb]
\centering
\includegraphics[width=0.49\textwidth]{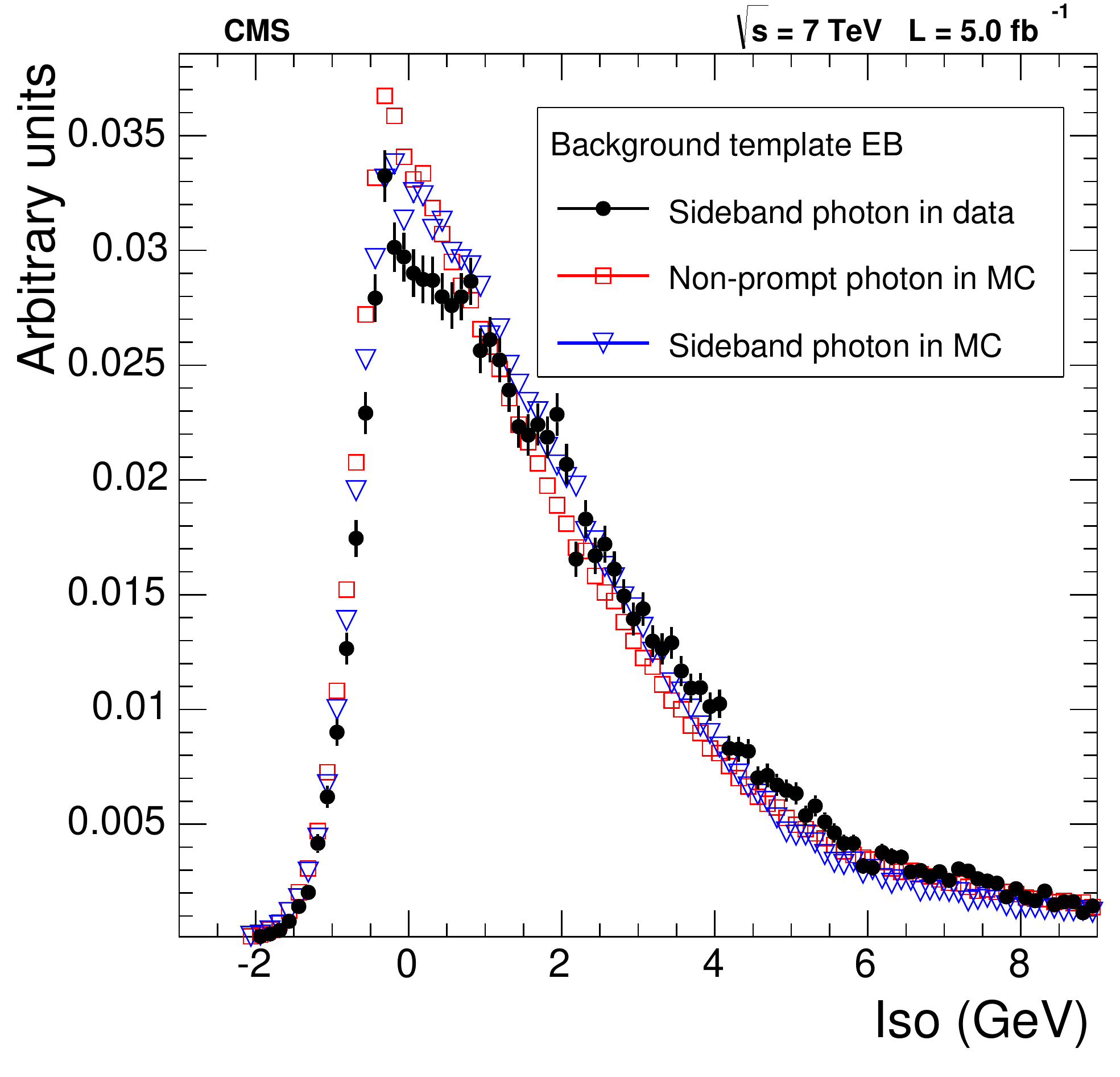} \hfill
\includegraphics[width=0.49\textwidth]{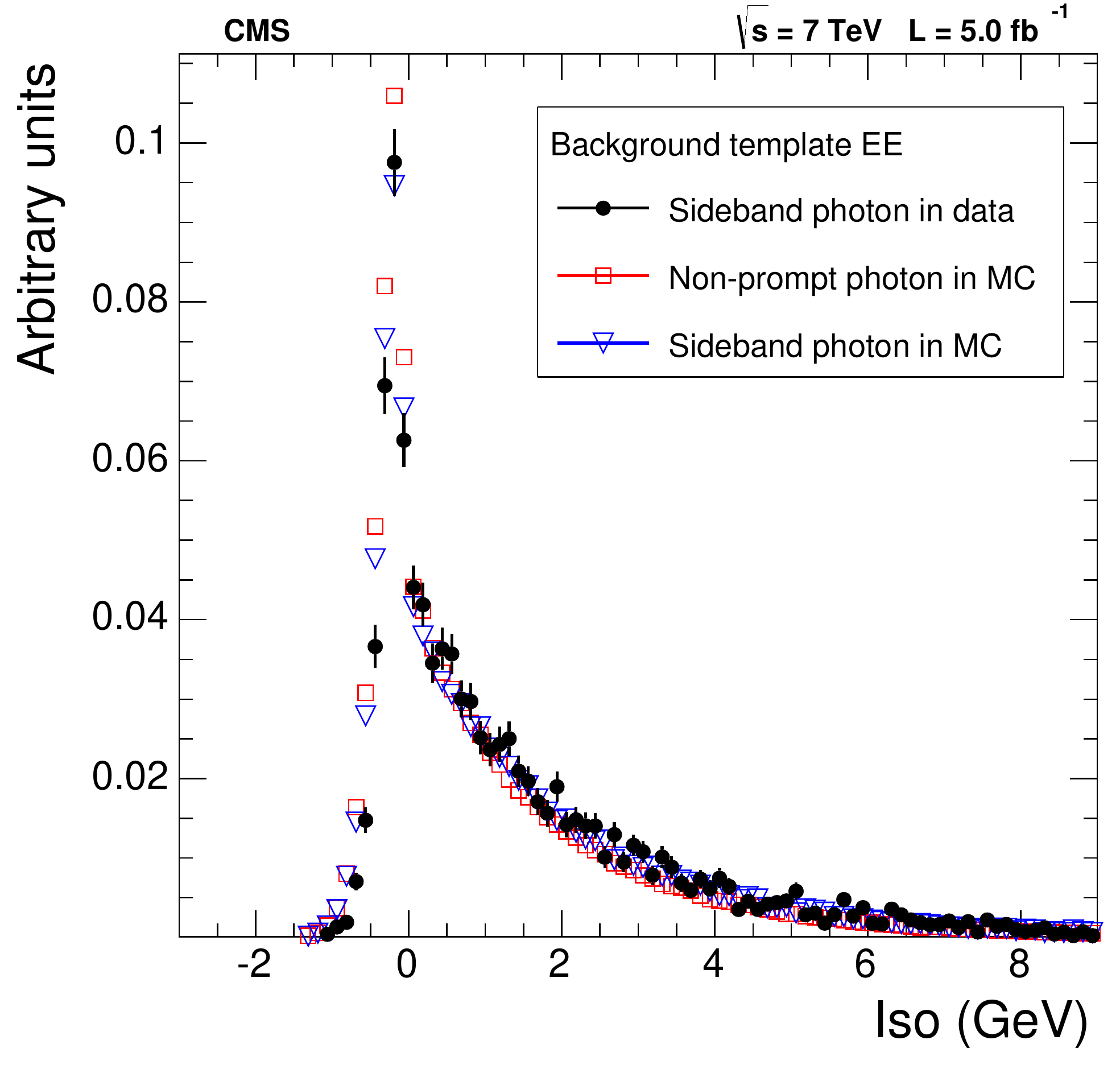}
\caption{Comparison of non-prompt photon templates in data and simulation: non-prompt photons in the simulation (squares), non-prompt photon templates extracted with the sideband technique from simulation (triangles) and from data (dots); (\cmsLeft) candidates in the ECAL barrel, (\cmsRight) candidates in the ECAL endcaps. All histograms are normalized to unit area.}
\label{fig:templateBkg}
\end{figure}

The same procedure (Section \ref{sec:PFIsolation}) is used for subtracting the pile-up energy from the photon isolation sums. The templates obtained using the random cone and the sideband techniques in the simulation are compared with the one-dimensional PF isolation distribution for prompt and non-prompt photons in simulated events and with the templates obtained from data (Figs.~\ref{fig:templateSignal} and \ref{fig:templateBkg}).

The residual differences in the simulation between the isolation distribution and the templates defined with the random cone and the sideband techniques are accounted for as systematic uncertainties on the template shapes.

The two-dimensional templates are built selecting candidate photons from data with the same kinematics as the diphoton events to be fitted. The procedure presented below correctly models the isolation distribution even in the case of overlap between the isolation cones of the two photon candidates.

The prompt-prompt template is built from events where the pileup energy density matches that of the event to be fitted, and where the two random cone directions are found having the same pseudorapidity and the same azimuthal angular separation as the selected photons.

The prompt-non-prompt template is built from events where a sideband photon is selected. The isolation sum around the sideband photon is used for the candidate to be fitted under the non-prompt-hypothesis. A direction satisfying the random cone criteria is then searched for in the same template event (oriented as the second candidate in the selected diphoton event) and used to calculate the isolation sum for the candidate to be fitted under the prompt hypothesis.

The non-prompt-non-prompt template is built selecting two events, each of which contains one sideband photon and such that their orientation matches the orientation of the candidate photons in the event to be fitted. Then, depending on the fraction of photon candidates with $\Delta R_{\gamma\gamma} < 1.0$ present in the bin of the observable under analysis, a choice between two different strategies is made.
If the fraction is below 10\%, the effect of the overlapping isolation cones can be neglected. The two-dimensional non-prompt-non-prompt template is then built by calculating each of the two isolation sums in the separate events. If the fraction is above 10\%, an additional requirement is imposed: the sum of the FastJet $\rho$ of the two selected template events has to match the one of the diphoton event to be fitted. Then, the sets of reconstructed particles in the two template events are merged, and the isolation sums are calculated from this merged set of reconstructed particles along the direction of each sideband photon.

In this procedure, the pileup energy density of the template events is used to model the pileup energy density of the event to be fitted, and this allows us to describe the correlation between the isolation sums. The effect of the residual correlation mis-modeling is added to the template shape systematic uncertainty in the final result.

\subsection{Fitting technique}

The fit is performed separately for the cases where both candidates are reconstructed in the ECAL barrel, one in the ECAL barrel and one in the ECAL endcaps, or both in the ECAL endcaps. If both candidates are in the same detector region (EB-EB and EE-EE categories), the leading selected photon is assigned randomly to axis 1 or 2 of the two-dimensional plane, and the prompt-non-prompt ($f_{pn}$) and non-prompt-prompt ($f_{np}$) fractions are constrained to have the same value.

{\tolerance=800
The fit, performed in each bin of the differential variables, is restricted to the region where the isolation of the photons is smaller than 9\GeV.
To guarantee its stability even in the less populated bins, the fit is performed in steps.
First the size of the bins in the two-dimensional plane $(\text{Iso}_1,\text{Iso}_2)$ is optimised to reduce statistical fluctuations of template shape in the tails; then a first fit is performed on the projections of the isolation distributions on the two axes of the plane using the one-dimensional templates described above.
In a subsequent step, the fractions of prompt-prompt, prompt-non-prompt, non-prompt-prompt, and non-prompt-non-prompt, which are constrained to sum up to unity, are fit in the two-dimensional plane using as a constraint the results of the previous fit. The final likelihood maximisation is then performed after removing all constraints, and using as initial values of the parameters those found in the previous step.
\par}

\begin{figure*}[htb]
\centering
\includegraphics[width=0.95\textwidth]{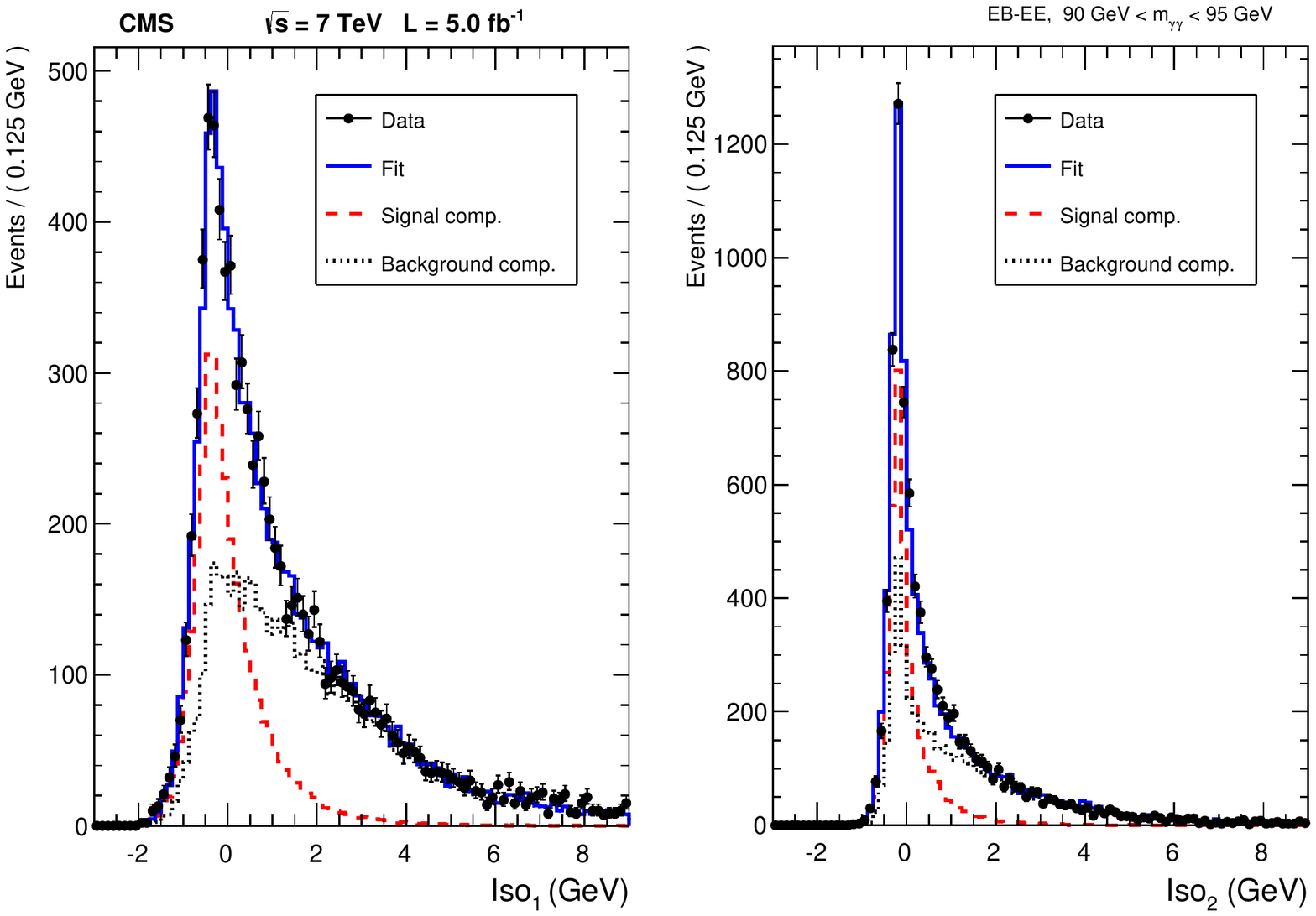}
\caption{Result of the first step of the fitting procedure, for the
$90\GeV<\mgg<95\GeV$
bin in the EB-EE category: isolation distribution for the photon reconstructed in the (left) ECAL barrel, (right) ECAL endcaps.}
\label{fig:projections}
\end{figure*}

\begin{figure*}[htb]
\centering
\includegraphics[width=0.95\textwidth]{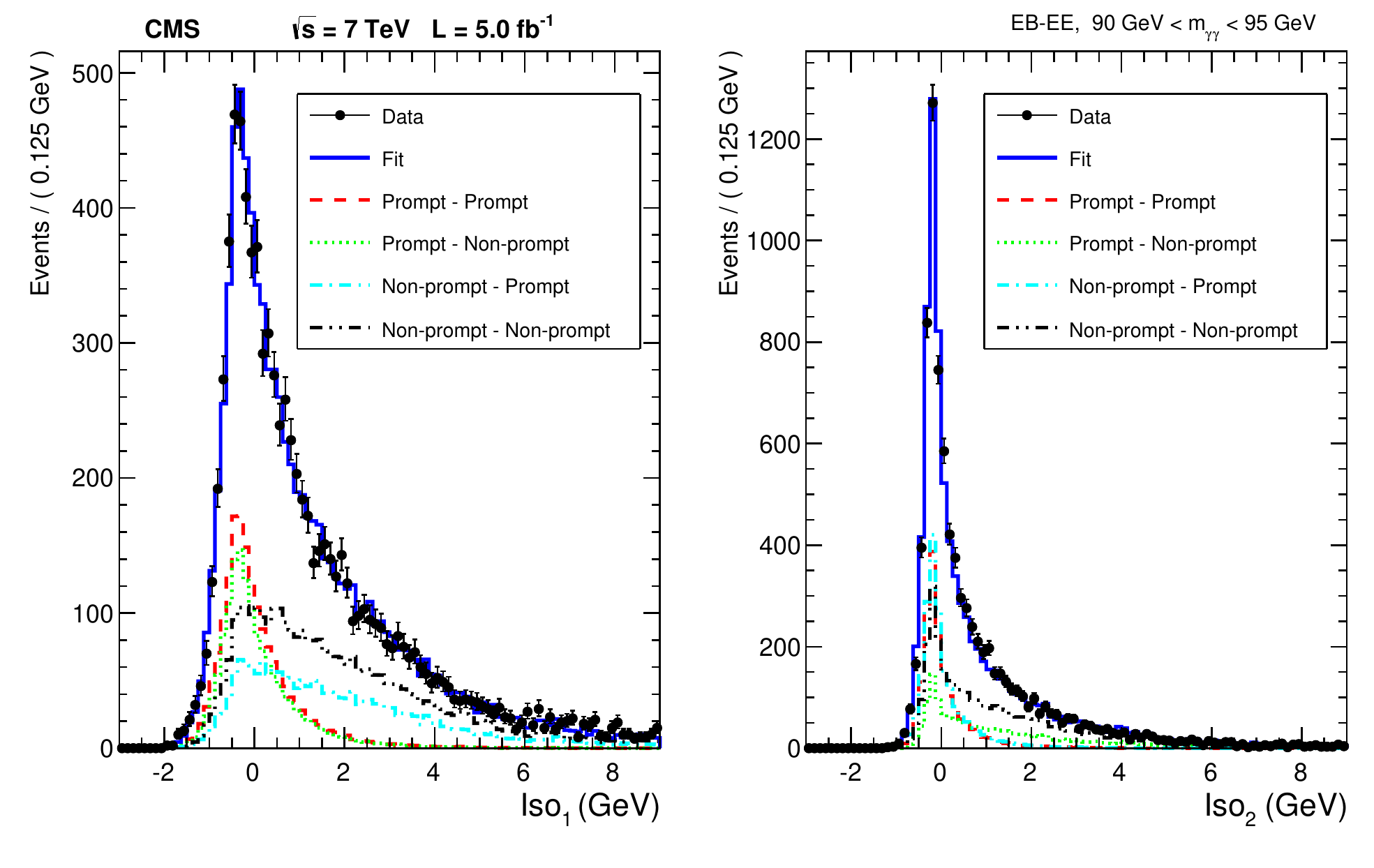}
\caption{Projections of the result of the final step of the fitting procedure, for the $90\GeV<\mgg<95\GeV$ bin in the EB-EE category: isolation distribution for the photon reconstructed in the (left) ECAL barrel, (right) ECAL endcaps.}
\label{fig:2Dfinalfit}
\end{figure*}

\begin{figure*}[htb]
\centering
\includegraphics[width=0.49\textwidth]{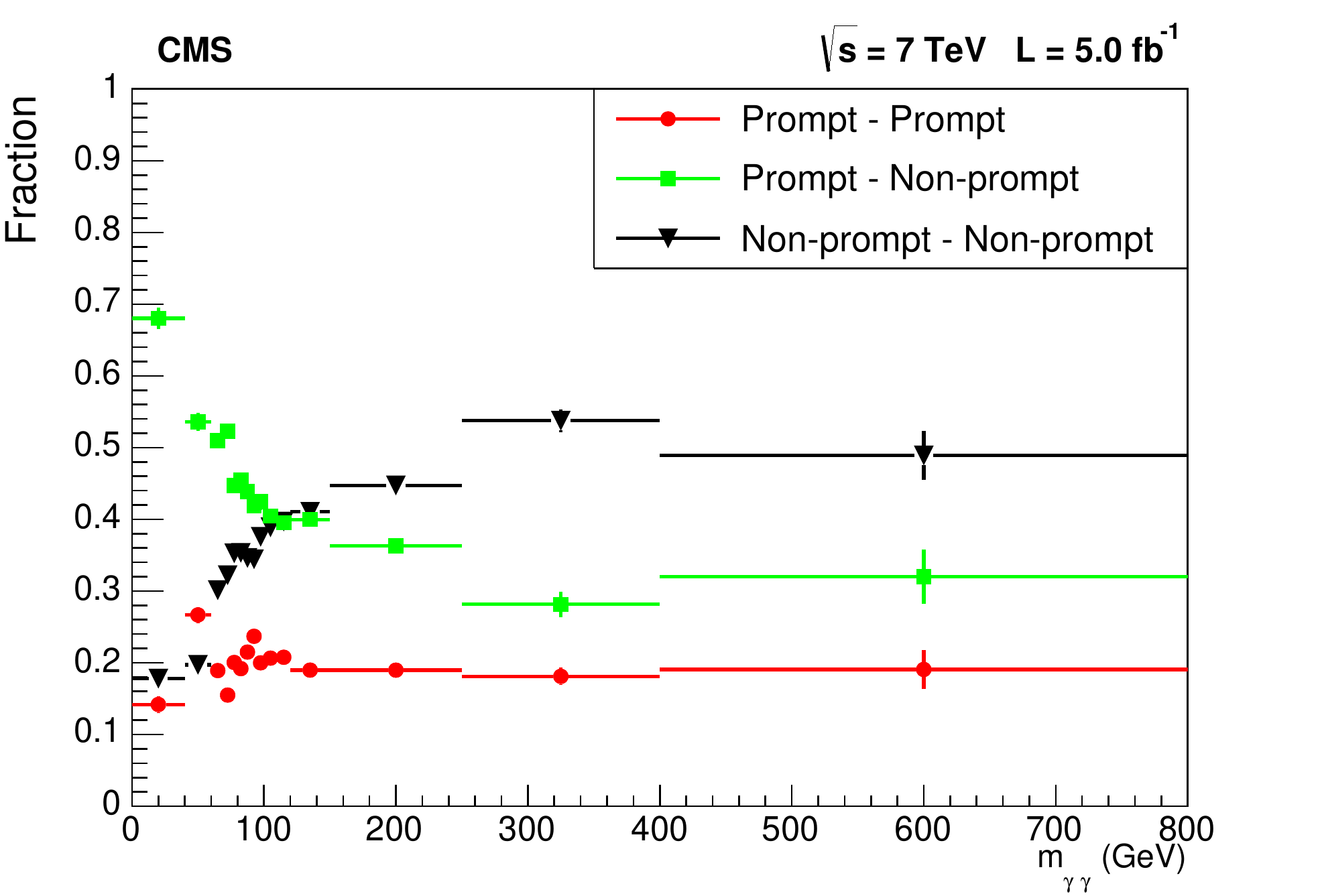}
\includegraphics[width=0.49\textwidth]{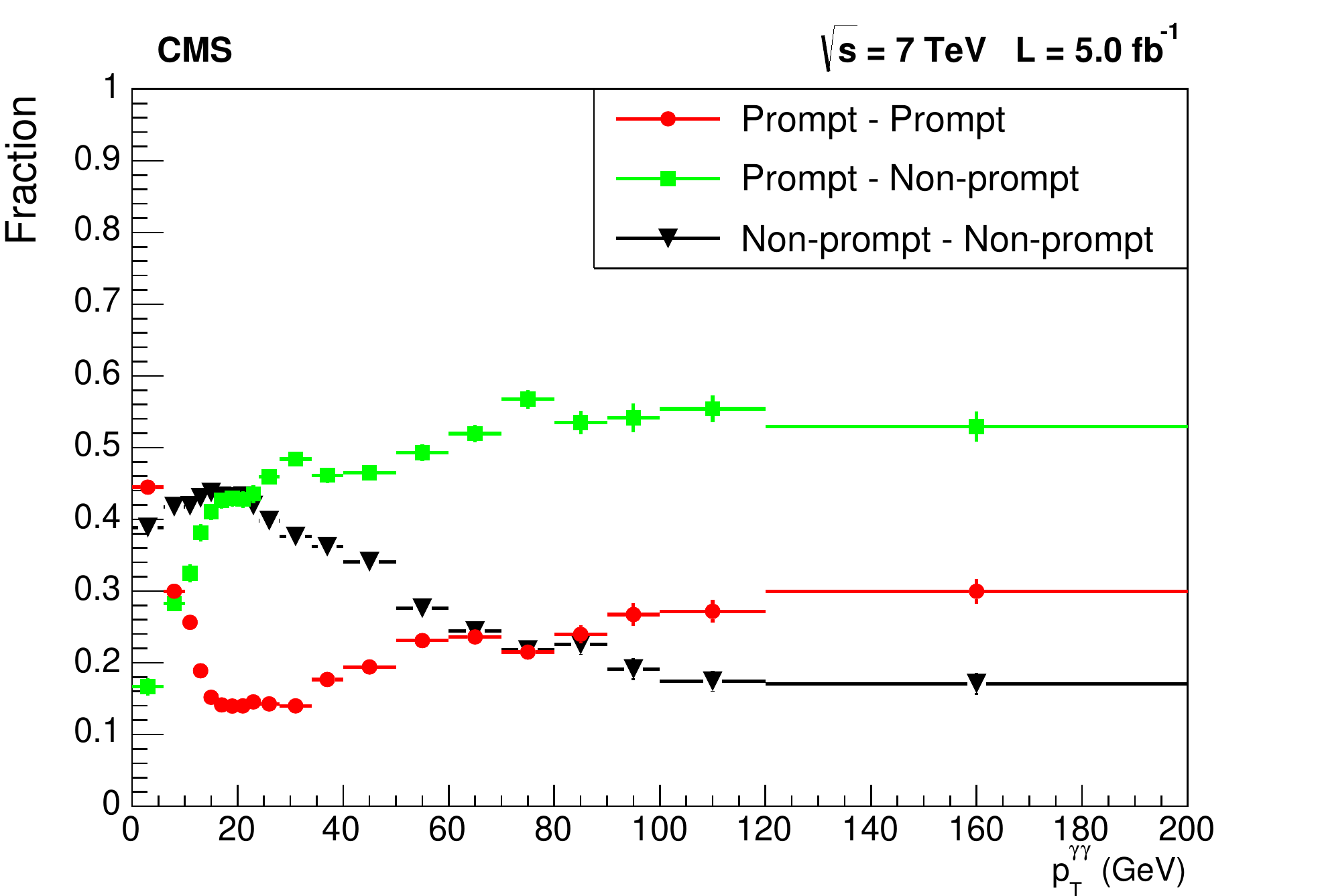} \includegraphics[width=0.49\textwidth]{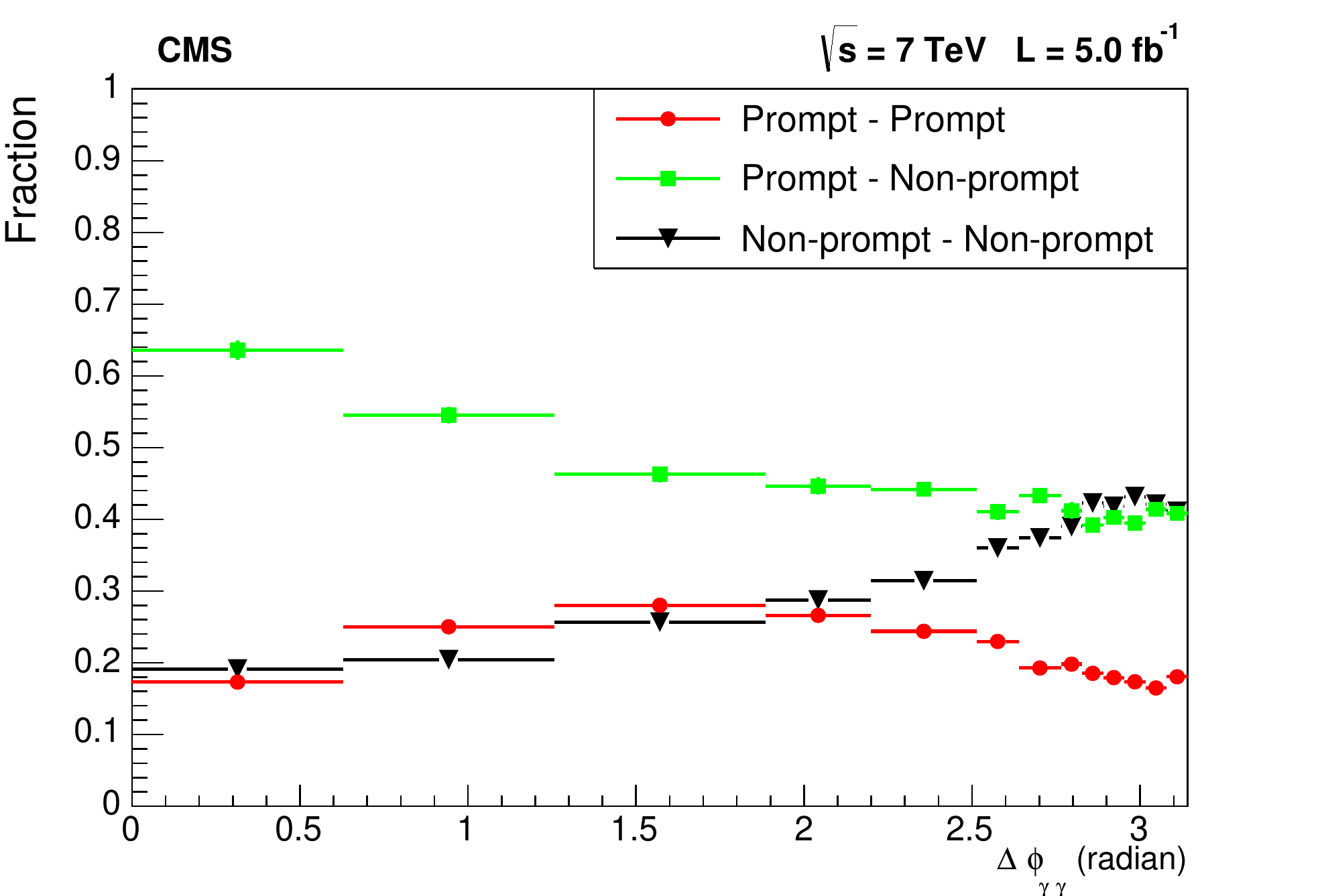}
\includegraphics[width=0.49\textwidth]{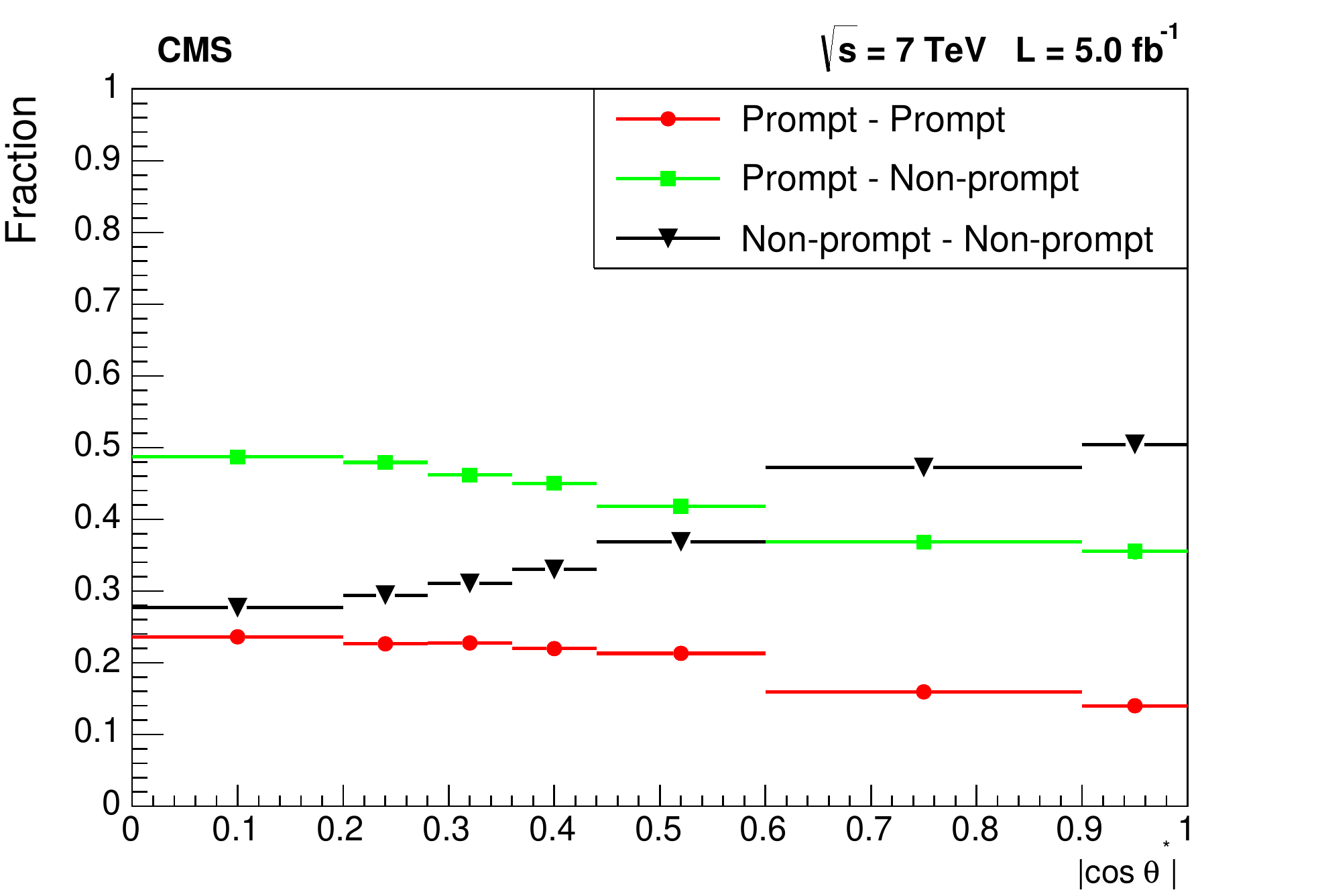}
\caption{Fractions of prompt-prompt, prompt-non-prompt and non-prompt-non-prompt components as a function of \mgg, \ptgg, \dphigg, $\abs{\cos\theta^*}$ in the whole acceptance of the analysis. Uncertainties are statistical only.}
\label{fig:differential}
\end{figure*}

An example of the first step of the procedure is obtained by fitting the one-dimensional projections of the isolation distributions as shown in Fig.~\ref{fig:projections}. An example of the results of the final two-dimensional fit (projected on the axes for the sake of clarity) is shown in Fig.~\ref{fig:2Dfinalfit}.
The fractions of prompt-prompt, prompt-non-prompt, and non-prompt-non-prompt components are shown in Fig.~\ref{fig:differential} for the observables of the differential analysis.
We fit about 69000 prompt diphoton events in the whole acceptance of the analysis.

The reported purity suffers from a contamination of electrons coming predominantly from Drell--Yan \Pep\Pem and incorrectly reconstructed as photons. The contamination is most significant in the \Z peak region, where it reaches about 25\% of the raw diphoton yield. The fraction of electron pairs passing the analysis selection and contributing to the prompt-prompt fitted fraction is estimated from simulation, where correction factors are applied to obtain the electron to photon mis-identification probability measured in data, and used to subtract the contamination.

\section{Efficiencies and unfolding}
\label{sec:Efficiencies}

Following the methodology presented in the previous sections, a ``raw'' diphoton production cross section is extracted. To obtain the final result, this cross section is corrected for inefficiencies and finally unfolded from the measured quantities to the corresponding particle-level quantities.

The total diphoton efficiency can be separated into the trigger efficiency and reconstruction/selection efficiency, and can be written as:
\ifthenelse{\boolean{cms@external}}{
\begin{multline*}
 \epsilon_{\Pgg\Pgg} =
 \epsilon_{\text{trig}} \times \epsilon_{\text{sel}}
 \times C_{\Pgg1}^{\Z \to \Pep\Pem} \times C_{\Pgg2}^{\Z \to \Pep\Pem}\\
 \times C_{\Pgg1}^{\Z \to \Pgmp\Pgmm\Pgg} \times C_{\Pgg2}^{\Z \to \Pgmp\Pgmm\Pgg}, \label{eq:diphotoneff}
\end{multline*}
}{
\begin{equation}
 \epsilon_{\Pgg\Pgg} = \epsilon_{\text{trig}} \times \epsilon_{\text{sel}}
 \times C_{\Pgg1}^{\Z \to \Pep\Pem} \times C_{\Pgg2}^{\Z \to \Pep\Pem}
 \times C_{\Pgg1}^{\Z \to \Pgmp\Pgmm\Pgg} \times C_{\Pgg2}^{\Z \to \Pgmp\Pgmm\Pgg}, \label{eq:diphotoneff}
\end{equation}
}
where $\epsilon_{\text{trig}}$ is the trigger efficiency and $\epsilon_{\text{sel}}$ is the diphoton
reconstruction/selection efficiency from simulation.
The factors $C_{\Pgg1}^{\Z \to \Pep\Pem}$ and $C_{\Pgg2}^{\Z \to \Pep\Pem}$ are
the corrections to the efficiency for each photon candidate to pass all
the selection requirements except the electron veto; $C_{\Pgg1}^{\Z \to \Pgmp\Pgmm\Pgg}$ and
$C_{\Pgg2}^{\Z \to \Pgmp\Pgmm\Pgg}$ are the corrections to the electron veto
efficiency.

The values of the correction factors are determined from the ratio of the efficiency in data to that in the simulation, measured with a tag-and-probe method using (i) samples of $\Z \to \Pep\Pem$ for the full selection except the electron-veto requirement, and (ii) samples of photons from the final-state-radiation of $\Z \to \Pgmp\Pgmm\Pgg$ for the electron-veto requirement.

The diphoton reconstruction/selection efficiency $\epsilon_{\text{sel}}$ is about 85\% when both photons are in the barrel, 75\% when one photon is in the barrel and the other in one endcap, and 64\% when both photons are in the endcaps. All these correction factors are estimated from data and range from 0.99 to 1.02, depending on the photon \ET and $\eta$.

The detector effects are unfolded from the measured yields for a direct comparison of experimental measurements
with theoretical predictions.
The number of unfolded diphoton events in each bin of the differential observables is obtained from
the reconstructed diphoton events in the data, $\vec{N}_\text{GEN}^\text{data}=M^{-1} \times \vec{N}_\text{RECO}^\text{data}$,
where the unfolding matrix $M$ is obtained from simulation, $\vec{N}_\text{RECO}^\mathrm{MC}=M \times \vec{N}_\text{GEN}^\mathrm{MC}$.
The unfolding matrix is calculated using the iterative Bayesian technique \cite{unfold:bayesian,RooUnfold}.
The diphoton simulated sample from \MADGRAPH hadronized with \PYTHIA is used. The distributions of diphoton candidates in the simulation are
reweighted to the distributions of the raw diphoton yields from data as obtained from the fit procedure,
for all the observables. The difference between the weighted and unweighted results
is taken into account as a systematic uncertainty, and amounts to about 1\%.
The unfolding correction amounts to 7\% of the raw yield at maximum, for the bins where the slope of the kinematic distributions is the steepest.

\section{Systematic uncertainties}
\label{sec:Systematics}

Table \ref{tab:systematics} summarises the main sources of systematic uncertainty in the measurement of the integrated cross section.

The dominant uncertainty in the template shapes arises from the difference in shape between the templates built with the techniques described in Section~\ref{sec:Purity} and the distributions of the isolation variable for prompt or non-prompt isolated photons for simulated events. The latter are used to generate data samples for each bin of the differential variables, with the fractions measured in data. Then, each of these datasets is fitted with templates built in the simulation with the same techniques used on data, and the average difference between the fitted fractions and those used for the generation is quoted as a systematic uncertainty. It amounts to 3\% (barrel template) and 5\% (endcap template) for the prompt component, and between 5\% (barrel template) and 10\% (endcap template) for the non-prompt component. The uncertainty in the template shape for fragmentation photons is evaluated in the simulation by doubling the probability of the fragmentation process, and that yields an additional 1.5\% uncertainty in the measured cross section. In the case of the non-prompt-non-prompt template, and only for the bins where a significant fraction of the diphoton candidates are close in $\Delta R_{\gamma\gamma}$, an additional uncertainty ranging from 3\% to 5\% is introduced to account for the imperfections on the template shape description due to the effect of ECAL noise and PF thresholds on the combination of two different events to build the template.

The systematic uncertainty arising from the statistical uncertainty in the shape of the templates is evaluated generating modified templates, where the content of each bin is represented by a Gaussian distribution centred on the nominal bin value and with standard deviation equal to the statistical uncertainty of the bin. The root mean square of the distribution of the fitted purity values, divided by the purity measured with the original template, is used as systematic uncertainty in the purity measurement and amounts to about 3\%.

A possible bias associated with the fitting procedure is evaluated using pseudo-experiments. Pseudo-data samples are generated with given fractions of prompt-prompt, prompt-non-prompt, and non-prompt-non-prompt contributions, using the templates from simulation as generator probability density functions. Each data sample is then fitted with the same templates used for the generation. The average bias is negligible in all bins.

The systematic uncertainty associated with the subtraction of Drell--Yan \Pep\Pem\ events is evaluated by propagating the uncertainty in the electron to photon misidentification probability to the subtracted yield. The uncertainty in the fraction of such events that is fitted as prompt-prompt is also taken into account. This contribution is maximal for \mgg close to the \Z-boson mass. The relative contribution to the total systematic uncertainty is below 0.5\%.

The systematic uncertainty in the trigger efficiency is found to be below 0.5\%. The systematic uncertainty in the reconstruction and selection efficiencies is dominated by the uncertainty in the data-to-simulation corrections from the $\Z \to \Pep\Pem$ and $\Z \to \Pgmp\Pgmm\Pgg$ control samples, and it ranges from 2\% in the barrel to 4\% in the endcap.

The systematic uncertainty in the integrated luminosity that corresponds to our data sample is 2.2\% \cite{CMS-PAS-SMP-12-008}.

The total systematic uncertainty in the measurement amounts to approximately 8\% when both candidates are reconstructed within the ECAL barrel, and to 11\% for the full acceptance of the analysis.

\begin{table}
\centering
\topcaption{Sources of systematic uncertainty in the measurement of integrated cross section.}
{
\begin{tabular}{lr}\hline
Source of uncertainty & \\ \hline
Prompt template shape (EB) & 3\% \\
Prompt template shape (EE) & 5\% \\
Non-prompt template shape (EB) & 5\% \\
Non-prompt template shape (EE) & 10\% \\
Effect of fragmentation component & 1.5\% \\
Template statistical fluctuation & 3\% \\
Selection efficiency & 2--4\% \\
Unfolding procedure & 1\% \\
Integrated luminosity & 2.2\% \\  \hline
\end{tabular}
}
\label{tab:systematics}
\end{table}

\section{Results and comparison with theoretical predictions}
\label{sec:Theory}

The measured unfolded differential cross sections are compared with the following generators for QCD diphoton production: \SHERPA 1.4.0 \cite{theory:Sherpa}, \DIPHOX 1.3.2 \cite{theory:Diphox} supplemented with \GENERATORGTOMC 1.1 \cite{theory:Gamma2MC}, \RESBOS \cite{theory:Resbos1, theory:Resbos2}, and \TWOGNNLO \cite{theory:2gNNLO}. Predictions with \SHERPA are computed at LO for the Born contribution with up to three additional real emissions (three extra jets) and with the box contribution at the matrix element level. The \DIPHOX NLO generator includes the direct and fragmentation contributions and uses a full fragmentation function for one or two partons into a photon at NLO. The direct box contribution, which is formally part of the NNLO corrections since it is initiated by gluon fusion through a quark loop, is computed at NLO with \GENERATORGTOMC. The \RESBOS NLO generator features resummation for Born and box contributions, and effectively includes fragmentation of one quark/gluon to a single photon at LO. The latter process is regulated to avoid divergences and does not include the full fragmentation function. The \RESBOS \ptgg spectrum benefits from a soft and collinear gluon resummation at next-to-next-to-leading-log accuracy. \TWOGNNLO predicts the direct $\Pgg\Pgg$+X processes at NNLO. The \SHERPA sample is used after hadronization while \DIPHOX + \GENERATORGTOMC, \RESBOS, and \TWOGNNLO are parton-level generators only and cannot be interfaced with parton shower generators.

The predictions have been computed for the phase space $\MYETONE>40$\GeV, $\MYETTWO>25$\GeV, $\abs{\eta_{\gamma}}<1.44$ or $1.57<\abs{\eta_{\gamma}}<2.5$, $\Delta R(\gamma_1,\gamma_2)>0.45$. An isolation requirement is applied at the generator level. In \SHERPA, the \ET sum of stable particles in a cone of size $\Delta R=0.4$ has to be less than 5\GeV (after hadronization). In \DIPHOX, \GENERATORGTOMC, and \RESBOS the \ET sum of partons in a cone of size $\Delta R=0.4$ is required to be less than 5\GeV. In \TWOGNNLO, the smooth Frixione isolation \cite{FrixioneIso} is applied to the photons to suppress the fragmentation component:

\begin{equation}
\ET^{\text{Iso}} (\Delta R) < \epsilon \left(  \frac{1-\cos(\Delta R)}{1-\cos(\Delta R_0)} \right)^{n},
\end{equation}

where $\ET^{\text{Iso}}$ is the \ET sum of partons in a cone of size $\Delta R$, $\Delta R_0=0.4$, $\epsilon=5$\GeV, and $n=0.05$. This criterion, tested with \DIPHOX, is found to have the same efficiency as that used for the other generators within a few percent. A non-perturbative correction is applied to \DIPHOX, \GENERATORGTOMC, and \TWOGNNLO predictions to correct for the fact that those generators do not include parton shower or underlying event contributions to the isolation cone.
The fraction of diphoton events not selected due to underlying hadronic activity falling inside the isolation cone is estimated using the \PYTHIA 6.4.22 \cite{Pythia6} event generator with tunes Z2, D6T, P0, and DWT \cite{UEpaper}. A factor of $0.95\pm 0.04$ is applied to the parton-level cross section to correct for this effect.

Theoretical predictions are performed using the CT10 \cite{CT10} NLO PDF set for \SHERPA, \DIPHOX + \GENERATORGTOMC, and \RESBOS, and the MSTW2008 \cite{MSTW2008} NNLO PDF set for \TWOGNNLO. The \DIPHOX and \GENERATORGTOMC theoretical uncertainties are computed in the following way: the factorization and renormalization scales in \GENERATORGTOMC are varied independently up and down by a factor of two around \mgg (configurations where one scale has a factor of four with respect to the other one are forbidden). In \DIPHOX, the factorization, renormalization and fragmentation scales are varied in the same way. In \RESBOS, the factorization and renormalization scales are varied simultaneously by a factor of two. The maximum and minimum values in each bin are used to define the uncertainty.
In \DIPHOX, \GENERATORGTOMC, and \RESBOS, the 52 CT10 eigenvector sets of PDFs are used to build the PDF uncertainty envelope,
also considering the uncertainty in the strong coupling constant \alpS, determined according to the CT10 \alpS PDF set. In \TWOGNNLO, a simplified and less computationally intensive estimate of the renormalization and factorization scale uncertainties is performed by varying these scales simultaneously by a factor of two up and down around \mgg; no PDF uncertainty is computed. The same procedure is used in \SHERPA, using the internal METS scale, where scales are defined as the lowest invariant mass or negative virtuality in the core 2$\to$2 configuration clustered using a \kt-type algorithm.

\begin{figure*}[htbp]
\centering
\includegraphics[width=0.45\textwidth]{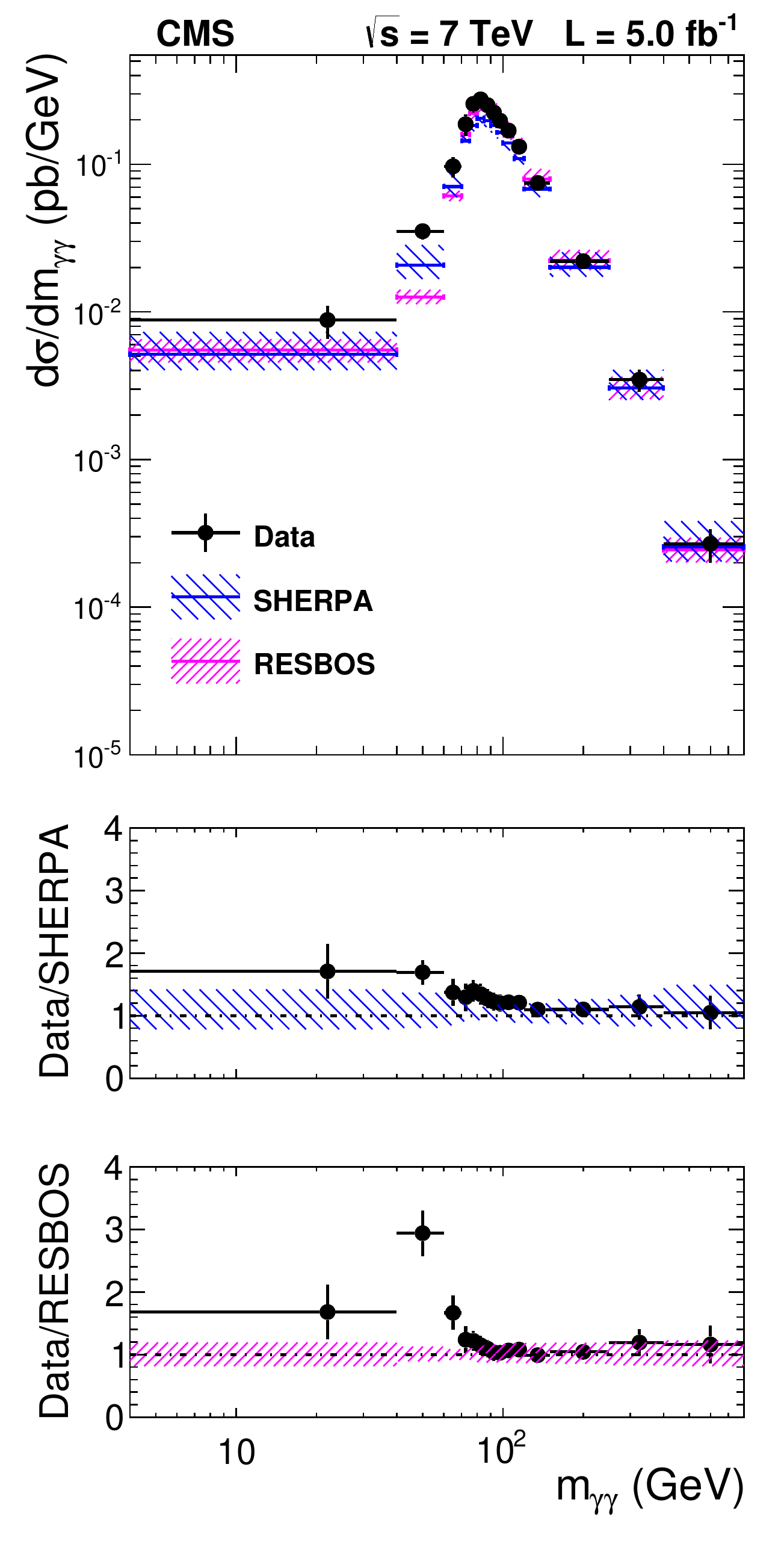}
\includegraphics[width=0.45\textwidth]{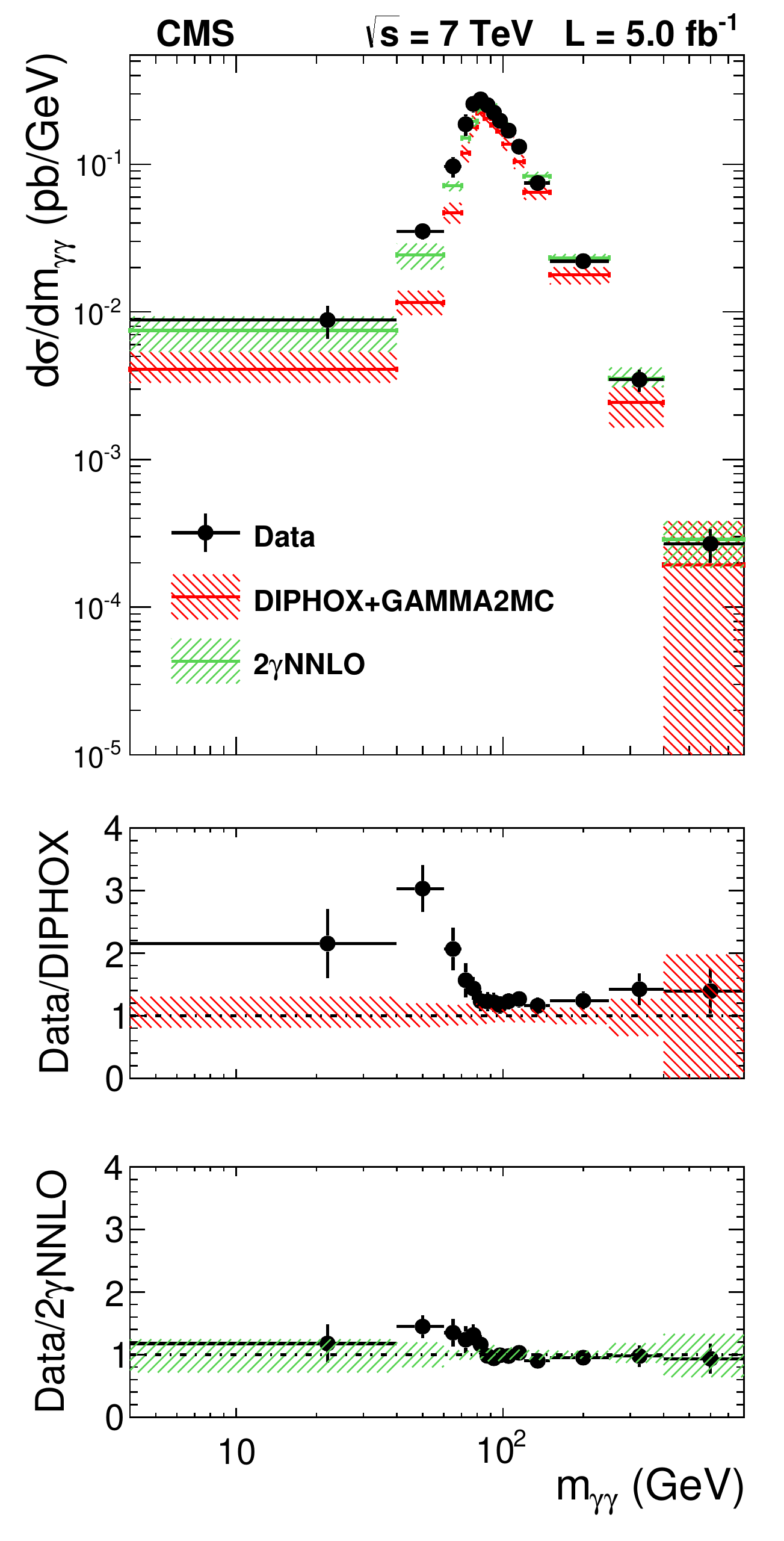}
\caption{ The comparisons of the differential cross section between
data and the \SHERPA, \DIPHOX + \GENERATORGTOMC, \RESBOS, and \TWOGNNLO predictions
for \mgg.
Black dots correspond to data with error bars including all statistical and systematic uncertainties.
Only the scale uncertainty is included for the \SHERPA prediction.
Scale, PDF and \alpS uncertainties are included for \DIPHOX + \GENERATORGTOMC and \RESBOS.
Only statistical and scale uncertainties are included for the \TWOGNNLO prediction.}
\label{fig:comparisonMASS}
\end{figure*}

\begin{figure*}[htbp]
\centering
\includegraphics[width=0.45\textwidth]{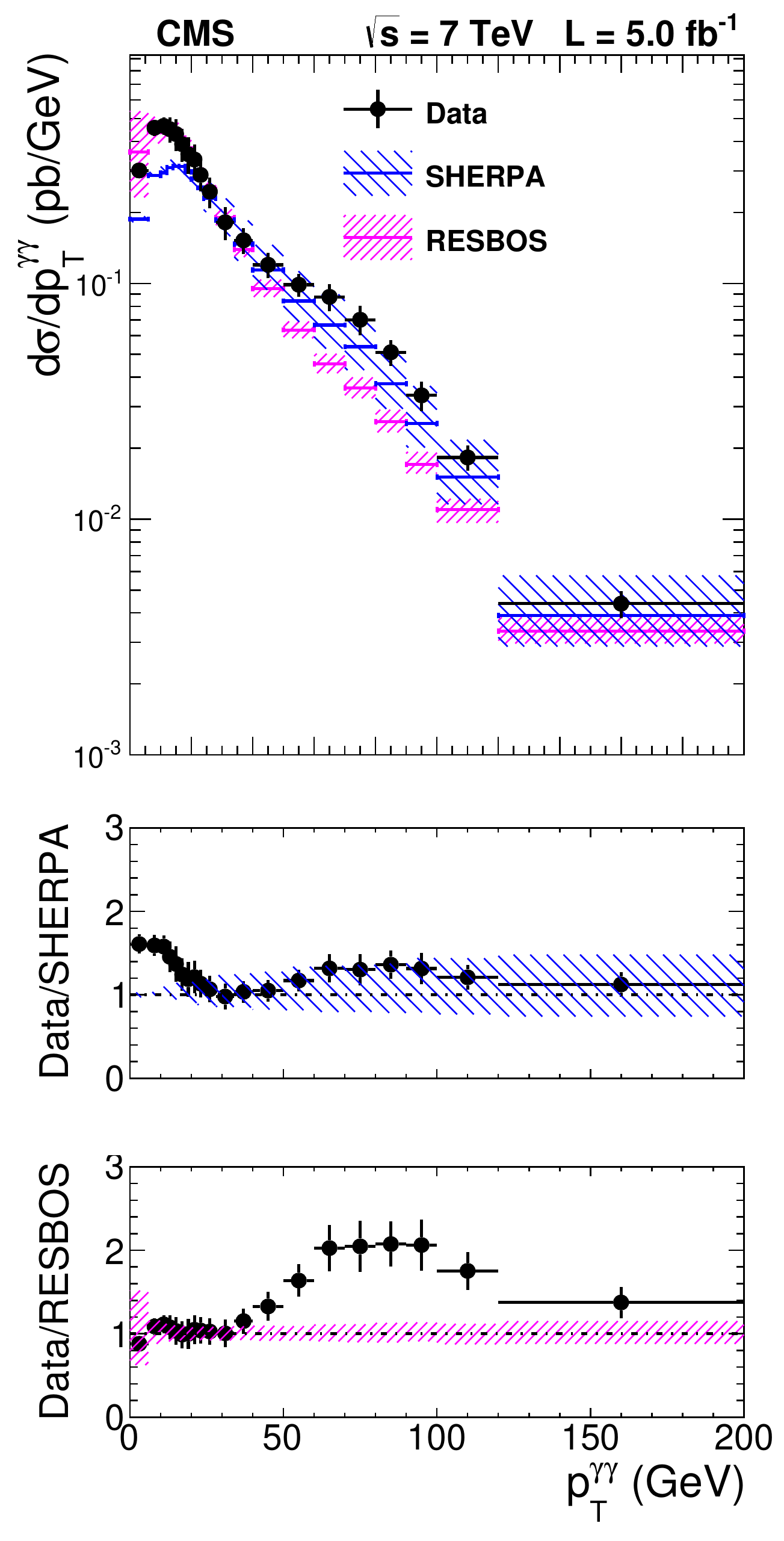}
\includegraphics[width=0.45\textwidth]{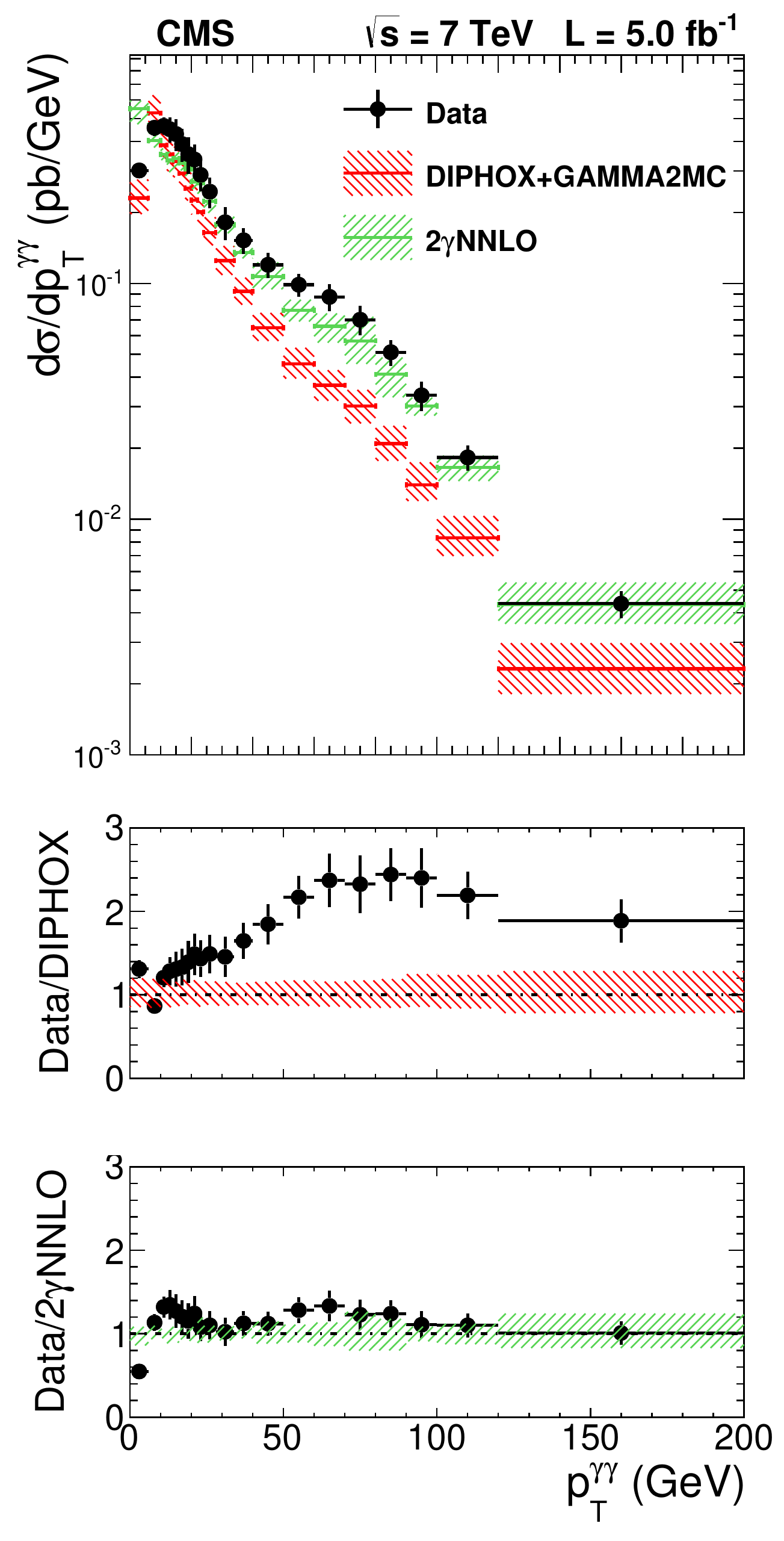}
\caption{ The comparisons of the differential cross section between
data and the \SHERPA, \DIPHOX + \GENERATORGTOMC, \RESBOS, and \TWOGNNLO predictions
for \ptgg.
Black dots correspond to data with error bars including all statistical and systematic uncertainties.
Only the scale uncertainty is included for the \SHERPA prediction.
Scale, PDF and \alpS uncertainties are included for \DIPHOX + \GENERATORGTOMC and \RESBOS.
Only statistical and scale uncertainties are included for the \TWOGNNLO prediction.}
\label{fig:comparisonQT}
\end{figure*}

\begin{figure*}[htbp]
\centering
\includegraphics[width=0.45\textwidth]{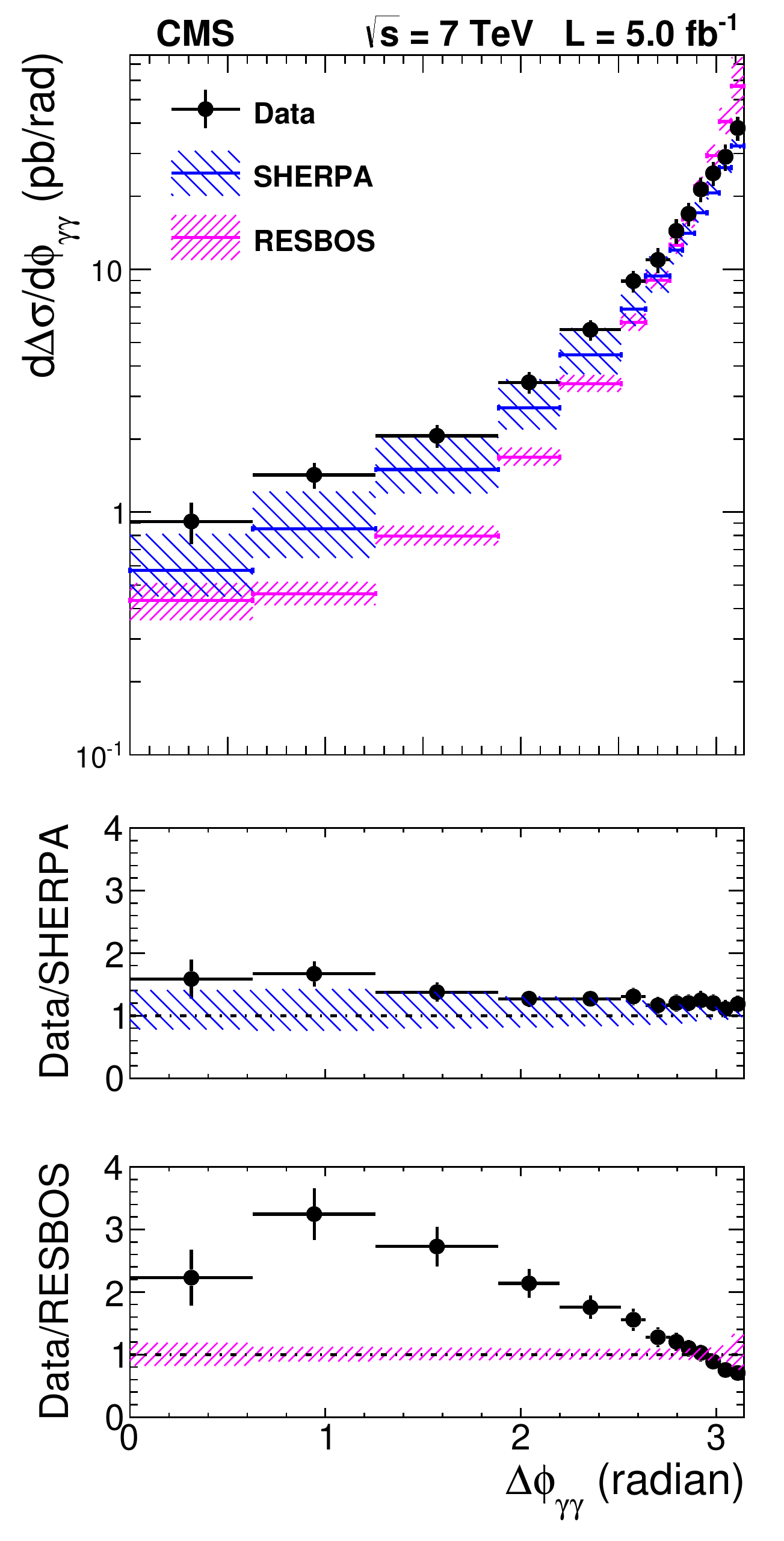}
\includegraphics[width=0.45\textwidth]{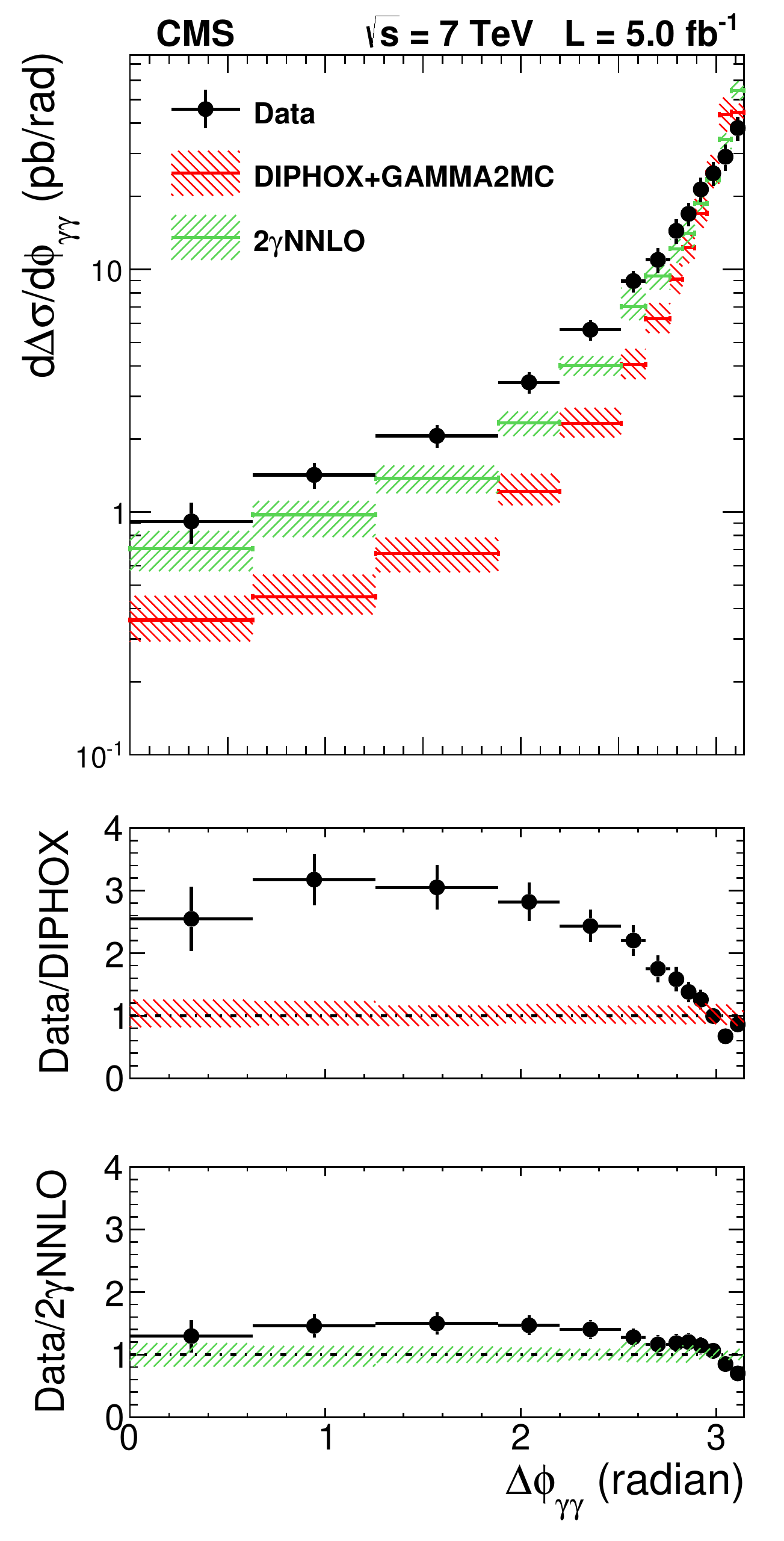}
\caption{ The comparisons of the differential cross section between
data and the \SHERPA, \DIPHOX + \GENERATORGTOMC, \RESBOS, and \TWOGNNLO predictions
for \dphigg.
Black dots correspond to data with error bars including all statistical and systematic uncertainties.
Only the scale uncertainty is included for the \SHERPA prediction.
Scale, PDF and \alpS uncertainties are included for \DIPHOX + \GENERATORGTOMC and \RESBOS.
Only statistical and scale uncertainties are included for the \TWOGNNLO prediction.}
\label{fig:comparisonDPHI}
\end{figure*}

\begin{figure*}[htbp]
\centering
\includegraphics[width=0.45\textwidth]{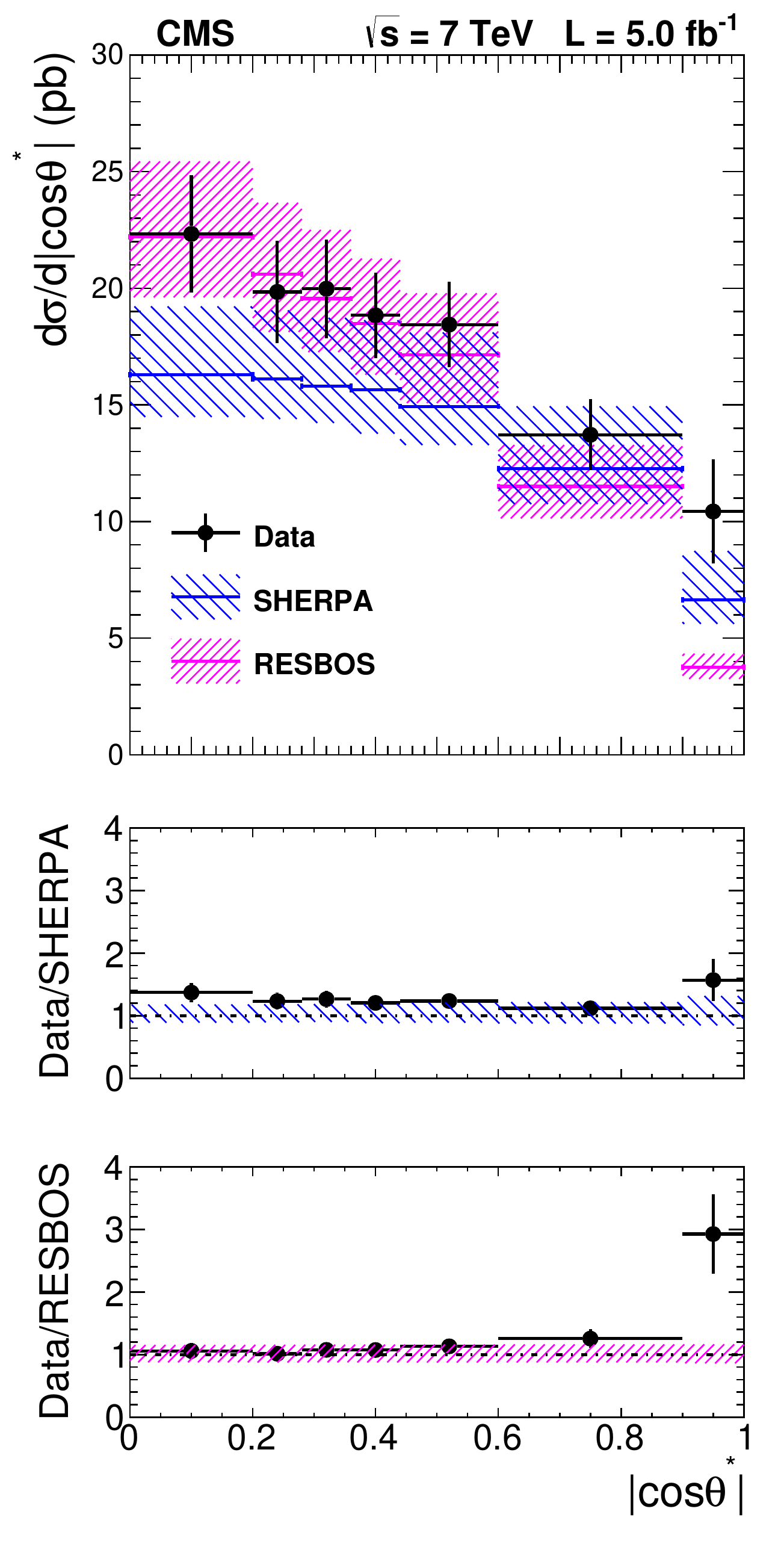}
\includegraphics[width=0.45\textwidth]{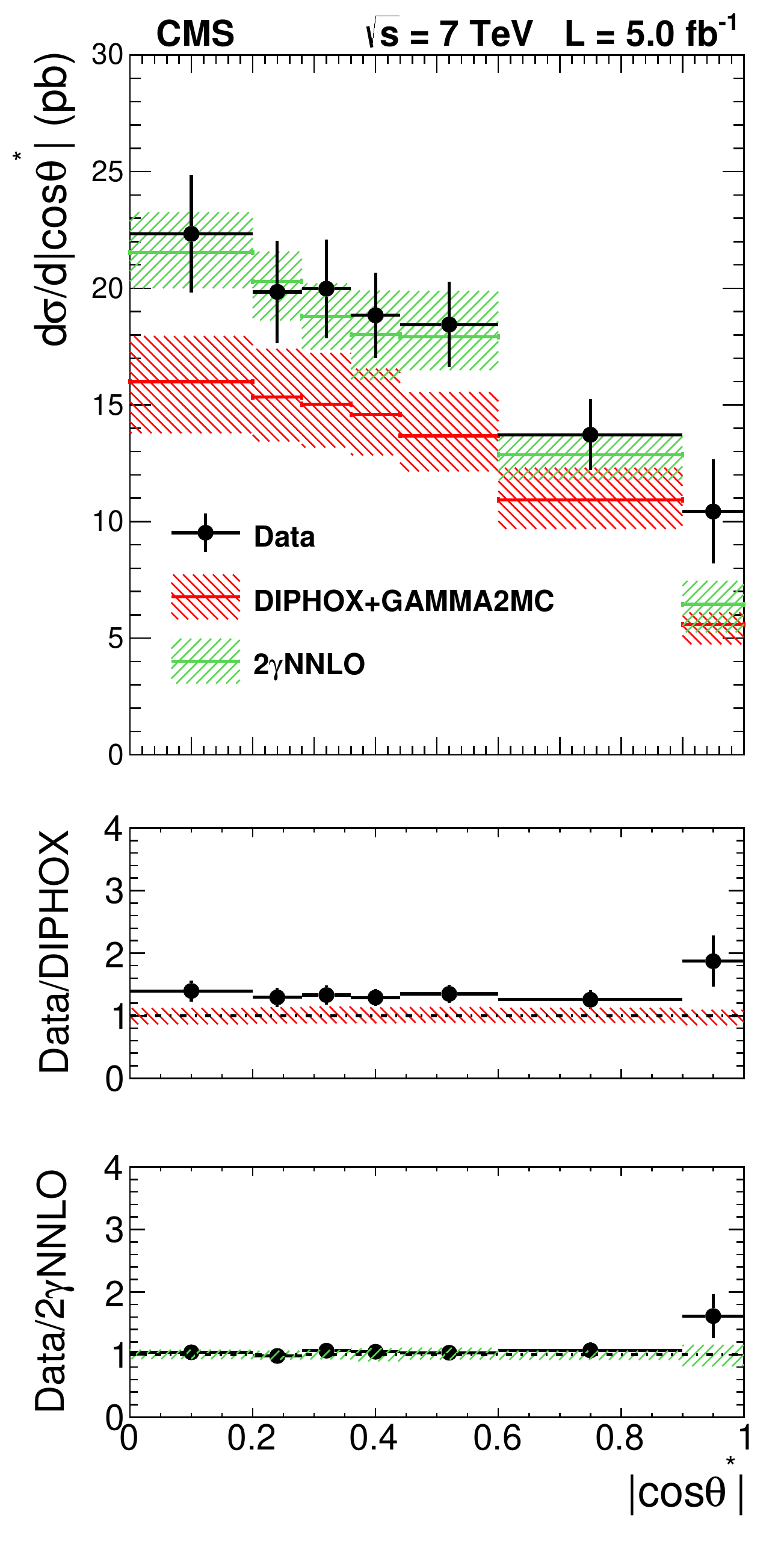} \caption{ The comparisons of the differential cross section between
data and the \SHERPA, \DIPHOX + \GENERATORGTOMC, \RESBOS, and \TWOGNNLO predictions
for $\abs{\cos\theta^*}$.
Black dots correspond to data with error bars including all statistical and systematic uncertainties.
Only the scale uncertainty is included for the \SHERPA prediction.
Scale, PDF and \alpS uncertainties are included for \DIPHOX + \GENERATORGTOMC and \RESBOS.
Only statistical and scale uncertainties are included for the \TWOGNNLO prediction.}
\label{fig:comparisonCOSTT}
\end{figure*}

The total cross section measured in data for the phase space defined above is:
\begin{equation*}
\sigma = 17.2 \pm 0.2\text{ (stat.)} \pm 1.9\text{ (syst.)} \pm 0.4\text{ (lum.)\unit{pb}},
\end{equation*}
compared with
\ifthenelse{\boolean{cms@external}}{
\begin{equation}\begin{aligned}
&\sigma_{\text{NNLO}}(\TWOGNNLO) = 16.2^{+1.5}_{-1.3}\,\text{(scale)}\unit{pb},\\
&\sigma_{\text{NLO}}(\DIPHOX+\GENERATORGTOMC) = \\
&\qquad 12.8^{+1.6}_{-1.5}\,\text{(scale)}^{+0.6}_{-0.8}\,\text{(pdf+}\alpS\text{)}\unit{pb},\\
&\sigma_{\text{NLO}}(\RESBOS) =14.9^{+2.2}_{-1.7}\,\text{(scale)} \pm 0.6\,\text{(pdf+}\alpS\text{)}\unit{pb},\\
&\sigma_{\text{LO}}(\SHERPA) =13.8^{+2.8}_{-1.6}\,\text{(scale)}\unit{pb}.
\end{aligned}\end{equation}
}{
\begin{equation*}
\begin{aligned}
\sigma_{\text{NNLO}}(\TWOGNNLO) &= 16.2^{+1.5}_{-1.3}\,\text{(scale)}\unit{pb},\\
\sigma_{\text{NLO}}(\DIPHOX+\GENERATORGTOMC) &= 12.8^{+1.6}_{-1.5}\,\text{(scale)}^{+0.6}_{-0.8}\,\text{(pdf+}\alpS\text{)}\unit{pb},\\
\sigma_{\text{NLO}}(\RESBOS) &= 14.9^{+2.2}_{-1.7}\,\text{(scale)} \pm 0.6\,\text{(pdf+}\alpS\text{)}\unit{pb},\\
\sigma_{\text{LO}}(\SHERPA) &= 13.8^{+2.8}_{-1.6}\,\text{(scale)}\unit{pb}.
\end{aligned}
\end{equation*}}
Figures \ref{fig:comparisonMASS}, \ref{fig:comparisonQT}, \ref{fig:comparisonDPHI} and  \ref{fig:comparisonCOSTT} show the comparisons of the differential cross section between data and the \SHERPA, \DIPHOX + \GENERATORGTOMC, \RESBOS, and \TWOGNNLO predictions for the four observables.

The NLO predictions of \DIPHOX + \GENERATORGTOMC are known to underestimate the data \cite{DiphotonPaper2010}, because of the missing higher-order contributions. Apart from an overall normalization factor, the phase space regions where the disagreement is the largest are at low \mgg, low \dphigg. The \RESBOS generator shows a similar trend, with a cross section closer to the data than \DIPHOX + \GENERATORGTOMC; its prediction is improved at high \dphigg due to soft gluon resummation. With higher-order diagrams included, \TWOGNNLO shows an improvement for the overall normalization. It also shows a better shape description, especially at low \dphigg, but it still underestimates the data in the same region. \SHERPA generally reproduces rather well the shape of the data, to a similar level as \TWOGNNLO. One can note that \TWOGNNLO and \SHERPA predict the \ptgg shoulder near $\MYETONE+\MYETTWO \sim 65$\GeV observed in the data. This is expected since \SHERPA includes up to three extra jets at the matrix element level.

\section{Summary}
\label{sec:Conclusions}

A measurement of differential cross sections for the production of a pair of isolated photons in \Pp\Pp\ collisions at $\sqrt{s} = 7$\TeV has been presented. The data sample corresponds to an integrated luminosity of 5.0\fbinv recorded in 2011 with the CMS detector. To enhance the sensitivity to higher-order diagrams, this measurement covers a phase space defined by an asymmetric \ET selection by requiring two isolated photons with \ET above 40 and 25\GeV respectively, in the pseudorapidity range $\abs{\eta}<2.5$, $\abs{\eta}\notin[1.44,1.57]$ and with an angular separation $\Delta R > 0.45$.

A data-driven method based on the photon component of the particle flow isolation has been used to extract the prompt diphoton yield. The isolation is calculated so that the energy leakage from the photon deposit inside the isolation cone is effectively subtracted.

The measured total cross section is
$$\sigma = 17.2 \pm 0.2\stat \pm 1.9\syst \pm 0.4\lum\unit{pb}$$
in agreement with the \TWOGNNLO prediction. The \SHERPA and \RESBOS predictions are compatible with the measurement within the uncertainties, while \DIPHOX + \GENERATORGTOMC underestimates the total cross section.

Differential cross sections for prompt diphoton production have been measured as a function of the diphoton invariant mass \mgg, the diphoton transverse momentum \ptgg, the azimuthal angular separation \dphigg between the two photons, and the cosine of the polar angle $\theta^{*}$ in the Collins--Soper frame of the diphoton system. The \TWOGNNLO and \SHERPA predictions show an improved agreement in shape with the data for the kinematic distributions with respect to the \DIPHOX + \GENERATORGTOMC and \RESBOS predictions, especially in the low $m_{\gamma\gamma}$, low $\Delta\phi_{\gamma\gamma}$ regions, which are the most sensitive to higher-order corrections.

\begin{acknowledgments}
We congratulate our colleagues in the CERN accelerator departments for the excellent performance of the LHC and thank the technical and administrative staffs at CERN and at other CMS institutes for their contributions to the success of the CMS effort. In addition, we gratefully acknowledge the computing centres and personnel of the Worldwide LHC Computing Grid for delivering so effectively the computing infrastructure essential to our analyses. Finally, we acknowledge the enduring support for the construction and operation of the LHC and the CMS detector provided by the following funding agencies: BMWFW and FWF (Austria); FNRS and FWO (Belgium); CNPq, CAPES, FAPERJ, and FAPESP (Brazil); MES (Bulgaria); CERN; CAS, MoST, and NSFC (China); COLCIENCIAS (Colombia); MSES and CSF (Croatia); RPF (Cyprus); MoER, ERC IUT and ERDF (Estonia); Academy of Finland, MEC, and HIP (Finland); CEA and CNRS/IN2P3 (France); BMBF, DFG, and HGF (Germany); GSRT (Greece); OTKA and NIH (Hungary); DAE and DST (India); IPM (Iran); SFI (Ireland); INFN (Italy); NRF and WCU (Republic of Korea); LAS (Lithuania); MOE and UM (Malaysia); CINVESTAV, CONACYT, SEP, and UASLP-FAI (Mexico); MBIE (New Zealand); PAEC (Pakistan); MSHE and NSC (Poland); FCT (Portugal); JINR (Dubna); MON, RosAtom, RAS and RFBR (Russia); MESTD (Serbia); SEIDI and CPAN (Spain); Swiss Funding Agencies (Switzerland); MST (Taipei); ThEPCenter, IPST, STAR and NSTDA (Thailand); TUBITAK and TAEK (Turkey); NASU and SFFR (Ukraine); STFC (United Kingdom); DOE and NSF (USA).

Individuals have received support from the Marie-Curie programme and the European Research Council and EPLANET (European Union); the Leventis Foundation; the A. P. Sloan Foundation; the Alexander von Humboldt Foundation; the Belgian Federal Science Policy Office; the Fonds pour la Formation \`a la Recherche dans l'Industrie et dans l'Agriculture (FRIA-Belgium); the Agentschap voor Innovatie door Wetenschap en Technologie (IWT-Belgium); the Ministry of Education, Youth and Sports (MEYS) of the Czech Republic; the Council of Science and Industrial Research, India; the Compagnia di San Paolo (Torino); and the Thalis and Aristeia programmes cofinanced by EU-ESF and the Greek NSRF.
\end{acknowledgments}

\vspace*{8em}
\bibliography{auto_generated}   

\providecommand{\href}[2]{#2}\begingroup\raggedright\begin{thebibliography}{10}%
\makeatletter
\providecommand{\hrefCMSnoop }[0]{\@secondoftwo}%
\makeatother
\providecommand{\doi}{\texttt{doi:}\begingroup \urlstyle{tt}\Url}

\bibitem{DiscoveryATLAS}
\hrefCMSnoop {} {{ {ATLAS}} Collaboration, ``Observation of a new particle in
  the search for the {Standard} {Model} {Higgs} boson with the {ATLAS} detector
  at the {LHC}'',} \textit{ Phys. Lett. B} \textbf{ 716} (2012) 1,
  \href{http://dx.doi.org/10.1016/j.physletb.2012.08.020}{\doi{10.1016/j.physletb.2012.08.020}},
\href{http://www.arXiv.org/abs/1207.7214}{\texttt{ arXiv:1207.7214}}.

\bibitem{DiscoveryCMS}
\hrefCMSnoop {} {{ {CMS}} Collaboration, ``Observation of a new boson at a mass
  of 125 {GeV} with the {CMS} experiment at the {LHC}'',} \textit{ Phys. Lett.
  B} \textbf{ 716} (2012) 30,
  \href{http://dx.doi.org/10.1016/j.physletb.2012.08.021}{\doi{10.1016/j.physletb.2012.08.021}},
\href{http://www.arXiv.org/abs/1207.7235}{\texttt{ arXiv:1207.7235}}.

\bibitem{CMSLongPaper}
\hrefCMSnoop {} {{ CMS} Collaboration, ``Observation of a new boson with mass
  near 125 {GeV} in pp collisions at $\sqrt{s}$ = 7 and 8 {TeV}'',} \textit{ J.
  High Ener. Phys.} \textbf{ 06} (2013) 081,
  \href{http://dx.doi.org/10.1007/JHEP06(2013)081}{\doi{10.1007/JHEP06(2013)081}},
\href{http://www.arXiv.org/abs/1303.4571}{\texttt{ arXiv:1303.4571}}.

\bibitem{GMSB}
\hrefCMSnoop {} {J.~T. Ruderman and D.~Shih, ``General neutralino {NLSPs} at
  the early {LHC}'',} \textit{ J. High Ener. Phys.} \textbf{ 08} (2012) 159,
  \href{http://dx.doi.org/10.1007/JHEP08(2012)159}{\doi{10.1007/JHEP08(2012)159}},
\href{http://www.arXiv.org/abs/1103.6083}{\texttt{ arXiv:1103.6083}}.

\bibitem{UED}
\hrefCMSnoop {} {N.~Arkani-Hamed, S.~Dimopoulos, and G.~R. Dvali, ``The
  hierarchy problem and new dimensions at a millimeter'',} \textit{ Phys. Lett.
  B} \textbf{ 429} (1998) 263,
  \href{http://dx.doi.org/10.1016/S0370-2693(98)00466-3}{\doi{10.1016/S0370-2693(98)00466-3}},
\href{http://www.arXiv.org/abs/hep-ph/9803315}{\texttt{ arXiv:hep-ph/9803315}}.

\bibitem{RS}
\hrefCMSnoop {} {L.~Randall and R.~Sundrum, ``A large mass hierarchy from a
  small extra dimension'',} \textit{ Phys. Rev. Lett.} \textbf{ 83} (1999)
  3370,
  \href{http://dx.doi.org/10.1103/PhysRevLett.83.3370}{\doi{10.1103/PhysRevLett.83.3370}},
\href{http://www.arXiv.org/abs/hep-ph/9905221}{\texttt{ arXiv:hep-ph/9905221}}.

\bibitem{RS2}
\hrefCMSnoop {} {L.~Randall and R.~Sundrum, ``{An alternative to
  compactification}'',} \textit{ Phys. Rev. Lett.} \textbf{ 83} (1999) 4690,
  \href{http://dx.doi.org/10.1103/PhysRevLett.83.4690}{\doi{10.1103/PhysRevLett.83.4690}},
\href{http://www.arXiv.org/abs/hep-th/9906064}{\texttt{ arXiv:hep-th/9906064}}.

\bibitem{diphotonCDF}
\hrefCMSnoop {} {{ CDF} Collaboration, ``Measurement of the cross section for
  prompt isolated diphoton production using the full {CDF} {Run} {II} data
  sample'',} \textit{ Phys. Rev. Lett.} \textbf{ 110} (2013) 101801,
  \href{http://dx.doi.org/10.1103/PhysRevLett.110.101801}{\doi{10.1103/PhysRevLett.110.101801}},
\href{http://www.arXiv.org/abs/1212.4204}{\texttt{ arXiv:1212.4204}}.

\bibitem{diphotonD0}
\hrefCMSnoop {} {{ D0} Collaboration, ``Measurement of the differential cross
  sections for isolated direct photon pair production in \Pp\Pap\ collisions at
  $\sqrt{s} = 1.96$ {TeV}'',} \textit{ Phys. Lett. B} \textbf{ 725} (2013) 6,
  \href{http://dx.doi.org/10.1016/j.physletb.2013.06.036}{\doi{10.1016/j.physletb.2013.06.036}},
\href{http://www.arXiv.org/abs/1301.4536}{\texttt{ arXiv:1301.4536}}.

\bibitem{diphotonATLAS}
\hrefCMSnoop {} {{ {ATLAS}} Collaboration, ``Measurement of isolated-photon
  pair production in \Pp\Pp\ collisions at $\sqrt{s}=7$ {TeV} with the {ATLAS}
  detector'',} \textit{ J. High Ener. Phys.} \textbf{ 01} (2013) 086,
  \href{http://dx.doi.org/10.1007/JHEP01(2013)086}{\doi{10.1007/JHEP01(2013)086}},
\href{http://www.arXiv.org/abs/1211.1913}{\texttt{ arXiv:1211.1913}}.

\bibitem{DiphotonPaper2010}
\hrefCMSnoop {} {{ {CMS}} Collaboration, ``Measurement of the production cross
  section for pairs of isolated photons in \Pp\Pp\ collisions at $\sqrt{s}=7$
  {TeV}'',} \textit{ J. High Ener. Phys.} \textbf{ 01} (2012) 133,
  \href{http://dx.doi.org/10.1007/JHEP01(2012)133}{\doi{10.1007/JHEP01(2012)133}},
\href{http://www.arXiv.org/abs/1110.6461}{\texttt{ arXiv:1110.6461}}.

\bibitem{QCD-10-019}
\hrefCMSnoop {} {{ {CMS}} Collaboration, ``Measurement of the isolated prompt
  photon production cross section in \Pp\Pp\ collisions at $\sqrt{s} =
  7$~{TeV}'',} \textit{ Phys. Rev. Lett.} \textbf{ 106} (2011) 082001,
  \href{http://dx.doi.org/10.1103/PhysRevLett.106.082001}{\doi{10.1103/PhysRevLett.106.082001}},
\href{http://www.arXiv.org/abs/1012.0799}{\texttt{ arXiv:1012.0799}}.

\bibitem{CMS-QCD-10-037}
\hrefCMSnoop {} {{ {CMS}} Collaboration, ``Measurement of the differential
  cross section for isolated prompt photon production in pp collisions at 7
  {TeV}'',} \textit{ Phys. Rev. D} \textbf{ 84} (2011) 052011,
  \href{http://dx.doi.org/10.1103/PhysRevD.84.052011}{\doi{10.1103/PhysRevD.84.052011}},
\href{http://www.arXiv.org/abs/1108.2044}{\texttt{ arXiv:1108.2044}}.

\bibitem{CMS-PAS-PFT-10-002}
\href {http://cdsweb.cern.ch/record/1279341} {{ CMS} Collaboration,
  ``Commissioning of the Particle-Flow reconstruction in Minimum-Bias and Jet
  Events from pp Collisions at 7 {TeV}'',} CMS Physics Analysis Summary
  CMS-PAS-PFT-10-002, 2010.

\bibitem{bib-detector}
\hrefCMSnoop {} {{ {CMS}} Collaboration, ``The {CMS} experiment at the {CERN}
  {LHC}'',} \textit{ J. Inst.} \textbf{ 3} (2008) S08004,
\href{http://dx.doi.org/10.1088/1748-0221/3/08/S08004}{\doi{10.1088/1748-0221/3/08/S08004}}.

\bibitem{WZpaper}
\hrefCMSnoop {} {{ CMS} Collaboration, ``Measurement of the inclusive W and Z
  production cross sections in pp collisions at $\sqrt{s}=7$ {TeV}'',} \textit{
  J. High Ener. Phys.} \textbf{ 10} (2011) 132,
  \href{http://dx.doi.org/10.1007/JHEP10(2011)132}{\doi{10.1007/JHEP10(2011)132}},
\href{http://www.arXiv.org/abs/1107.4789}{\texttt{ arXiv:1107.4789}}.

\bibitem{Madgraph5}
J.~Alwall\hrefCMSnoop {} { {et~al.}, ``{MadGraph 5: Going Beyond}'',} \textit{
  J. High Ener. Phys.} \textbf{ 06} (2011) 128,
  \href{http://dx.doi.org/10.1007/JHEP06(2011)128}{\doi{10.1007/JHEP06(2011)128}},
\href{http://www.arXiv.org/abs/1106.0522}{\texttt{ arXiv:1106.0522}}.

\bibitem{Pythia6}
\hrefCMSnoop {} {T.~Sj{\"o}strand, S.~Mrenna, and P.~Z. Skands, ``{PYTHIA} 6.4
  Physics and Manual'',} \textit{ J. High Ener. Phys.} \textbf{ 05} (2006) 026,
  \href{http://dx.doi.org/10.1088/1126-6708/2006/05/026}{\doi{10.1088/1126-6708/2006/05/026}},
\href{http://www.arXiv.org/abs/hep-ph/0603175}{\texttt{ arXiv:hep-ph/0603175}}.

\bibitem{CTEQ6}
J.~Pumplin\hrefCMSnoop {} { {et~al.}, ``{New generation of parton distributions
  with uncertainties from global QCD analysis}'',} \textit{ J. High Ener.
  Phys.} \textbf{ 07} (2002) 012,
  \href{http://dx.doi.org/10.1088/1126-6708/2002/07/012}{\doi{10.1088/1126-6708/2002/07/012}},
\href{http://www.arXiv.org/abs/hep-ph/0201195}{\texttt{ arXiv:hep-ph/0201195}}.

\bibitem{UEpaper}
\hrefCMSnoop {} {{ CMS} Collaboration, ``Measurement of the underlying event
  activity at the {LHC} with $\sqrt{s}= 7$ {TeV} and comparison with $\sqrt{s}
  = 0.9$ {TeV}'',} \textit{ J. High Ener. Phys.} \textbf{ 09} (2011) 109,
  \href{http://dx.doi.org/10.1007/JHEP09(2011)109}{\doi{10.1007/JHEP09(2011)109}},
\href{http://www.arXiv.org/abs/1107.0330}{\texttt{ arXiv:1107.0330}}.

\bibitem{Agostinelli:2002hh}
\hrefCMSnoop {} {{ GEANT4} Collaboration, ``{GEANT4}: A simulation toolkit'',}
  \textit{ Nucl. Instrum. Meth. A} \textbf{ 506} (2003) 250,
  \href{http://dx.doi.org/10.1016/S0168-9002(03)01368-8}{\doi{10.1016/S0168-9002(03)01368-8}}.

\bibitem{EGM-11-001}
\hrefCMSnoop {} {{ CMS} Collaboration, ``Energy calibration and resolution of
  the {CMS} electromagnetic calorimeter in pp collisions at $\sqrt{s}=7$
  {TeV}'',} \textit{ J. Inst.} \textbf{ 8} (2013) P09009,
  \href{http://dx.doi.org/10.1088/1748-0221/8/09/P09009}{\doi{10.1088/1748-0221/8/09/P09009}},
\href{http://www.arXiv.org/abs/1306.2016}{\texttt{ arXiv:1306.2016}}.

\bibitem{2005_gsf_paper}
\hrefCMSnoop {} {W.~Adam, R.~Fr{\"u}hwirth, A.~Strandlie, and T.~Todorov,
  ``Reconstruction of electrons with the Gaussian-sum filter in the CMS tracker
  at the LHC'',} \textit{ J. Phys. G} \textbf{ 31} (2005) N9,
  \href{http://dx.doi.org/10.1088/0954-3899/31/9/N01}{\doi{10.1088/0954-3899/31/9/N01}}.

\bibitem{CollinsSoper}
\hrefCMSnoop {} {J.~C. Collins and D.~E. Soper, ``Angular distribution of
  dileptons in high-energy hadron collisions'',} \textit{ Phys. Rev. D}
  \textbf{ 16} (1977) 2219,
  \href{http://dx.doi.org/10.1103/PhysRevD.16.2219}{\doi{10.1103/PhysRevD.16.2219}}.

\bibitem{fastjet}
\hrefCMSnoop {} {M.~Cacciari and G.~P. Salam, ``Pileup subtraction using jet
  areas'',} \textit{ Phys. Lett. B} \textbf{ 659} (2008) 119,
  \href{http://dx.doi.org/10.1016/j.physletb.2007.09.077}{\doi{10.1016/j.physletb.2007.09.077}},
  \href{http://www.arXiv.org/abs/0707.1378}{\texttt{ arXiv:0707.1378}}.

\bibitem{unfold:bayesian}
\hrefCMSnoop {} {G.~D'Agostini, ``A multidimensional unfolding method based on
  Bayes theorem'',} \textit{ Nucl. Instrum. Meth. A} \textbf{ 372} (1996) 46,
\href{http://dx.doi.org/10.1016/0168-9002(95)00274-X}{\doi{10.1016/0168-9002(95)00274-X}}.

\bibitem{RooUnfold}
\hrefCMSnoop {} {T.~Adye, ``{Unfolding algorithms and tests using
  RooUnfold}'',} in \textit{ Proceedings of the PHYSTAT 2011 Workshop on
  Statistical Issues Related to Discovery Claims in Search Experiments and
  Unfolding}, p.~313.
\newblock 2011.
\newblock
  \href{http://dx.doi.org/10.5170/CERN-2011-006}{\doi{10.5170/CERN-2011-006}}.

\bibitem{CMS-PAS-SMP-12-008}
\href {http://cdsweb.cern.ch/record/1434360} {{ CMS} Collaboration, ``Absolute
  calibration of the luminosity measurement at {CMS}: {W}inter 2012 update'',}
  CMS Physics Analysis Summary CMS-PAS-SMP-12-008, 2012.

\bibitem{theory:Sherpa}
T.~Gleisberg\hrefCMSnoop {} { {et~al.}, ``Event generation with {SHERPA}
  1.1'',} \textit{ J. High Ener. Phys.} \textbf{ 02} (2009) 007,
  \href{http://dx.doi.org/10.1088/1126-6708/2009/02/007}{\doi{10.1088/1126-6708/2009/02/007}},
\href{http://www.arXiv.org/abs/0811.4622}{\texttt{ arXiv:0811.4622}}.

\bibitem{theory:Diphox}
\hrefCMSnoop {} {T.~Binoth, J.~P. Guillet, E.~Pilon, and M.~Werlen, ``A full
  next-to-leading order study of direct photon pair production in hadronic
  collisions'',} \textit{ Eur. Phys. J. C} \textbf{ 16} (2000) 311,
  \href{http://dx.doi.org/10.1007/s100520050024}{\doi{10.1007/s100520050024}},
\href{http://www.arXiv.org/abs/hep-ph/9911340}{\texttt{ arXiv:hep-ph/9911340}}.

\bibitem{theory:Gamma2MC}
\hrefCMSnoop {} {Z.~Bern, L.~J. Dixon, and C.~Schmidt, ``Isolating a light
  {Higgs} boson from the diphoton background at the {CERN} {LHC}'',} \textit{
  Phys. Rev. D} \textbf{ 66} (2002) 074018,
  \href{http://dx.doi.org/10.1103/PhysRevD.66.074018}{\doi{10.1103/PhysRevD.66.074018}},
\href{http://www.arXiv.org/abs/hep-ph/0206194}{\texttt{ arXiv:hep-ph/0206194}}.

\bibitem{theory:Resbos1}
\hrefCMSnoop {} {C.~Balazs, E.~L. Berger, S.~Mrenna, and C.~P. Yuan, ``Photon
  pair production with soft gluon resummation in hadronic interactions'',}
  \textit{ Phys. Rev. D} \textbf{ 57} (1998) 6934,
  \href{http://dx.doi.org/10.1103/PhysRevD.57.6934}{\doi{10.1103/PhysRevD.57.6934}}.

\bibitem{theory:Resbos2}
\hrefCMSnoop {} {C.~Bal{\'a}zs, E.~L. Berger, P.~M. Nadolsky, and C.-P. Yuan,
  ``Calculation of prompt diphoton production cross-sections at {Tevatron} and
  {LHC} energies'',} \textit{ Phys. Rev. D} \textbf{ 76} (2007) 013009,
  \href{http://dx.doi.org/10.1103/PhysRevD.76.013009}{\doi{10.1103/PhysRevD.76.013009}}.

\bibitem{theory:2gNNLO}
S.~Catani\hrefCMSnoop {} { {et~al.}, ``Diphoton production at hadron colliders:
  a fully-differential {QCD} calculation at {NNLO}'',} \textit{ Phys. Rev.
  Lett.} \textbf{ 108} (2012) 072001,
  \href{http://dx.doi.org/10.1103/PhysRevLett.108.072001}{\doi{10.1103/PhysRevLett.108.072001}},
\href{http://www.arXiv.org/abs/1110.2375}{\texttt{ arXiv:1110.2375}}.

\bibitem{FrixioneIso}
\hrefCMSnoop {} {S.~Frixione, ``Isolated photons in perturbative {QCD}'',}
  \textit{ Phys. Lett. B} \textbf{ 429} (1998) 369,
  \href{http://dx.doi.org/10.1016/S0370-2693(98)00454-7}{\doi{10.1016/S0370-2693(98)00454-7}},
\href{http://www.arXiv.org/abs/hep-ph/9801442}{\texttt{ arXiv:hep-ph/9801442}}.

\bibitem{CT10}
H.-L. Lai\hrefCMSnoop {} { {et~al.}, ``New parton distributions for collider
  physics'',} \textit{ Phys. Rev. D} \textbf{ 82} (2010) 074024,
  \href{http://dx.doi.org/10.1103/PhysRevD.82.074024}{\doi{10.1103/PhysRevD.82.074024}},
\href{http://www.arXiv.org/abs/1007.2241}{\texttt{ arXiv:1007.2241}}.

\bibitem{MSTW2008}
\hrefCMSnoop {} {A.~D. Martin, W.~J. Stirling, R.~S. Thorne, and G.~Watt,
  ``Parton distributions for the {LHC}'',} \textit{ Eur. Phys. J. C} \textbf{
  63} (2009) 189,
  \href{http://dx.doi.org/10.1140/epjc/s10052-009-1072-5}{\doi{10.1140/epjc/s10052-009-1072-5}},
\href{http://www.arXiv.org/abs/0901.0002}{\texttt{ arXiv:0901.0002}}.

\end{thebibliography}\endgroup
\appendix
\section{Cross section tables}
\label{sec:tables}

The numerical values of the cross sections for each bin of the diphoton invariant mass \mgg, the diphoton transverse momentum \ptgg, the azimuthal angle difference \dphigg between the two photons, and the cosine of the polar angle $\theta^{*}$ in the Collins--Soper frame of the diphoton pair  are presented in Tables \ref{tab:2gNNLOPredictionMgg}, \ref{tab:2gNNLOPredictionQt}, \ref{tab:2gNNLOPredictionDphi}, and \ref{tab:2gNNLOPredictionCostt}.

\renewcommand{\arraystretch}{1.2}
\begin{table*}[htb]
\centering
\topcaption{Values of $\rd\sigma/\rd{}m_{\Pgg\Pgg}$ (pb/\GeVns) for the data, \SHERPA, \DIPHOX+\GENERATORGTOMC, \RESBOS and \TWOGNNLO predictions.}
{\footnotesize

\begin{tabular}{c|ccccc}
\hline
     &  \multicolumn{5}{c}{ $\rd\sigma/\rd{}m_{\Pgg\Pgg}$ (pb/\GeVns{})} \\
\hline
\mgg& Data  &  \SHERPA & \DIPHOX & \RESBOS  &  \TWOGNNLO \\
(\GeVns)&   &   & +\GENERATORGTOMC & & \\
\hline
  \multirow{2}{*}{0--40}& 0.0088 & 0.0052 & 0.0041 $^{+29\%}_{-17\%}$(scale)  & 0.0052 $^{+8.4\%}_{-6.1\%}$(scale)  & 0.0075 $^{+23\%}_{-27\%}$(scale) \\
 &  $\pm$ 25\%(tot.) & $^{+42\%}_{-22\%}$(scale) &  \hspace{6mm} $^{+8.0\%}_{-6.8\%}$(pdf) & \hspace{0mm} $\pm$17\%(pdf+$\alpha_{s}$)  &  \hspace{4mm}  $\pm$ 6.1\%(stat.) \\
\hline
  \multirow{2}{*}{40--60}& 0.035 & 0.021 & 0.012 $^{+19\%}_{-16\%}$(scale)  & 0.012 $^{+10\%}_{-7.6\%}$(scale)  & 0.024 $^{+18\%}_{-19\%}$(scale) \\
 &  $\pm$ 11\%(tot.) & $^{+37\%}_{-20\%}$(scale) &  \hspace{6mm} $^{+3.8\%}_{-6.2\%}$(pdf) & \hspace{0mm} $\pm$6.7\%(pdf+$\alpha_{s}$)  &  \hspace{4mm}  $\pm$ 3.5\%(stat.) \\
\hline
  \multirow{2}{*}{60--70}& 0.097 & 0.071 & 0.047 $^{+15\%}_{-14\%}$(scale)  & 0.058 $^{+6.5\%}_{-6.3\%}$(scale)  & 0.072 $^{+4.3\%}_{-6.5\%}$(scale) \\
 &  $\pm$ 15\%(tot.) & $^{+23\%}_{-15\%}$(scale) &  \hspace{6mm} $^{+4.7\%}_{-3.8\%}$(pdf) & \hspace{0mm} $\pm$3.8\%(pdf+$\alpha_{s}$)  &  \hspace{4mm}  $\pm$ 3.5\%(stat.) \\
\hline
  \multirow{2}{*}{70--75}& 0.187 & 0.144 & 0.119 $^{+17\%}_{-12\%}$(scale)  & 0.151 $^{+6.9\%}_{-5.5\%}$(scale)  & 0.151 $^{+3.8\%}_{-4.2\%}$(scale) \\
 &  $\pm$ 17\%(tot.) & $^{+20\%}_{-9.1\%}$(scale) &  \hspace{6mm} $^{+6.1\%}_{-1.9\%}$(pdf) & \hspace{0mm} $\pm$4.1\%(pdf+$\alpha_{s}$)  &  \hspace{4mm}  $\pm$ 2.3\%(stat.) \\
\hline
  \multirow{2}{*}{75--80}& 0.256 & 0.183 & 0.179 $^{+18\%}_{-13\%}$(scale)  & 0.210 $^{+7.7\%}_{-7.1\%}$(scale)  & 0.195 $^{+5.0\%}_{-6.7\%}$(scale) \\
 &  $\pm$ 12\%(tot.) & $^{+17\%}_{-8.8\%}$(scale) &  \hspace{6mm} $^{+2.1\%}_{-4.5\%}$(pdf) & \hspace{0mm} $\pm$4.1\%(pdf+$\alpha_{s}$)  &  \hspace{4mm}  $\pm$ 1.9\%(stat.) \\
\hline
  \multirow{2}{*}{80--85}& 0.275 & 0.204 & 0.223 $^{+14\%}_{-12\%}$(scale)  & 0.239 $^{+12\%}_{-10\%}$(scale)  & 0.237 $^{+5.1\%}_{-1.2\%}$(scale) \\
 &  $\pm$ 12\%(tot.) & $^{+14\%}_{-7.1\%}$(scale) &  \hspace{6mm} $^{+0.2\%}_{-7.7\%}$(pdf) & \hspace{0mm} $\pm$4.1\%(pdf+$\alpha_{s}$)  &  \hspace{4mm}  $\pm$ 5.4\%(stat.) \\
\hline
  \multirow{2}{*}{85--90}& 0.251 & 0.198 & 0.205 $^{+11\%}_{-11\%}$(scale)  & 0.230 $^{+14\%}_{-11\%}$(scale)  & 0.257 $^{+14\%}_{-9.6\%}$(scale) \\
 &  $\pm$ 12\%(tot.) & $^{+15\%}_{-9.5\%}$(scale) &  \hspace{6mm} $^{+1.7\%}_{-4.4\%}$(pdf) & \hspace{0mm} $\pm$3.9\%(pdf+$\alpha_{s}$)  &  \hspace{4mm}  $\pm$ 2.8\%(stat.) \\
\hline
  \multirow{2}{*}{90--95}& 0.224 & 0.183 & 0.184 $^{+12\%}_{-9.2\%}$(scale)  & 0.217 $^{+15\%}_{-11\%}$(scale)  & 0.238 $^{+14\%}_{-7.2\%}$(scale) \\
 &  $\pm$ 11\%(tot.) & $^{+14\%}_{-10\%}$(scale) &  \hspace{6mm} $^{+2.8\%}_{-3.1\%}$(pdf) & \hspace{0mm} $\pm$3.7\%(pdf+$\alpha_{s}$)  &  \hspace{4mm}  $\pm$ 4.2\%(stat.) \\
\hline
  \multirow{2}{*}{95--100}& 0.197 & 0.164 & 0.169 $^{+9.3\%}_{-11\%}$(scale)  & 0.192 $^{+15\%}_{-11\%}$(scale)  & 0.199 $^{+6.0\%}_{-8.0\%}$(scale) \\
 &  $\pm$ 11\%(tot.) & $^{+18\%}_{-8.8\%}$(scale) &  \hspace{6mm} $^{+-0.4\%}_{-5.6\%}$(pdf) & \hspace{0mm} $\pm$4.3\%(pdf+$\alpha_{s}$)  &  \hspace{4mm}  $\pm$ 4.7\%(stat.) \\
\hline
  \multirow{2}{*}{100--110}& 0.169 & 0.139 & 0.137 $^{+9.5\%}_{-8.5\%}$(scale)  & 0.159 $^{+16\%}_{-11\%}$(scale)  & 0.173 $^{+10\%}_{-5.7\%}$(scale) \\
 &  $\pm$ 9.4\%(tot.) & $^{+16\%}_{-11\%}$(scale) &  \hspace{6mm} $^{+2.1\%}_{-3.5\%}$(pdf) & \hspace{0mm} $\pm$4.1\%(pdf+$\alpha_{s}$)  &  \hspace{4mm}  $\pm$ 3.6\%(stat.) \\
\hline
  \multirow{2}{*}{110--120}& 0.131 & 0.109 & 0.103 $^{+9.1\%}_{-9.0\%}$(scale)  & 0.122 $^{+17\%}_{-12\%}$(scale)  & 0.128 $^{+5.0\%}_{-3.9\%}$(scale) \\
 &  $\pm$ 9.1\%(tot.) & $^{+17\%}_{-11\%}$(scale) &  \hspace{6mm} $^{+4.69\%}_{-3.1\%}$(pdf) & \hspace{0mm} $\pm$3.6\%(pdf+$\alpha_{s}$)  &  \hspace{4mm}  $\pm$ 4.7\%(stat.) \\
\hline
  \multirow{2}{*}{120--150}& 0.075 & 0.068 & 0.064 $^{+8.8\%}_{-8.7\%}$(scale)  & 0.075 $^{+17\%}_{-12\%}$(scale)  & 0.083 $^{+3.4\%}_{-5.9\%}$(scale) \\
 &  $\pm$ 10\%(tot.) & $^{+20\%}_{-12\%}$(scale) &  \hspace{6mm} $^{+3.2\%}_{-4.7\%}$(pdf) & \hspace{0mm} $\pm$3.6\%(pdf+$\alpha_{s}$)  &  \hspace{4mm}  $\pm$ 4.6\%(stat.) \\
\hline
 \multirow{2}{*}{150--250}& 0.022 & 0.020 & 0.018 $^{+10\%}_{-9.8\%}$(scale)  & 0.021 $^{+18\%}_{-13\%}$(scale)  & 0.023 $^{+3.3\%}_{-4.0\%}$(scale) \\
 &  $\pm$ 10\%(tot.) & $^{+26\%}_{-14\%}$(scale) &  \hspace{6mm} $^{+6.8\%}_{-8.3\%}$(pdf) & \hspace{0mm} $\pm$3.6\%(pdf+$\alpha_{s}$)  &  \hspace{4mm}  $\pm$ 3.2\%(stat.) \\
\hline
  \multirow{2}{*}{250--400}& 0.0035 & 0.0030 & 0.0024 $^{+18\%}_{-21\%}$(scale)  & 0.0029 $^{+18\%}_{-16\%}$(scale)  & 0.0036 $^{+16\%}_{-10\%}$(scale) \\
 &  $\pm$ 17\%(tot.) & $^{+33\%}_{-17\%}$(scale) &  \hspace{6mm} $^{+19\%}_{-24\%}$(pdf) & \hspace{0mm} $\pm$3.8\%(pdf+$\alpha_{s}$)  &  \hspace{4mm}  $\pm$ 7.2\%(stat.) \\
\hline
  \multirow{2}{*}{400--800}& 0.0003 & 0.0003 & 0.0002 $^{+68\%}_{-71\%}$(scale)  & 0.0002 $^{+21\%}_{-17\%}$(scale)  & 0.0003 $^{+21\%}_{-25\%}$(scale) \\
 &  $\pm$ 25\%(tot.) & $^{+49\%}_{-21\%}$(scale) &  \hspace{6mm} $^{+69\%}_{-75\%}$(pdf) & \hspace{0mm} $\pm$4.6\%(pdf+$\alpha_{s}$)  &  \hspace{4mm}  $\pm$ 25\%(stat.) \\
\hline
\end{tabular}
}
\label{tab:2gNNLOPredictionMgg}
\end{table*}

\begin{table*}[htbp]
\centering
\topcaption{Values of $\rd\sigma/\rd\pt^{\Pgg\Pgg}$ (pb/\GeVns) for the data, \SHERPA, \DIPHOX+\GENERATORGTOMC, \RESBOS and \TWOGNNLO predictions.}
{\footnotesize
\begin{tabular}{c|ccccc}
\hline
     &  \multicolumn{5}{c}{ $\rd\sigma/\rd\pt^{\Pgg\Pgg}$ (pb/\GeVns{})} \\
\hline
\ptgg& Data  &  \SHERPA & \DIPHOX & \RESBOS  &  \TWOGNNLO \\
(\GeVns)&   &   & +\GENERATORGTOMC & & \\
\hline
  \multirow{2}{*}{0--6}& 0.302 & 0.187 & 0.230 $^{+17\%}_{-13\%}$(scale)  & 0.343 $^{+51\%}_{-37\%}$(scale)  & 0.551 $^{+7.0\%}_{-13\%}$(scale) \\
 &  $\pm$ 7.2\%(tot.) & $^{+2.8\%}_{-3.0\%}$(scale) &  \hspace{6mm} $^{+8.3\%}_{-2.0\%}$(pdf) & \hspace{0mm} $\pm$4.7\%(pdf+$\alpha_{s}$)  &  \hspace{4mm}  $\pm$ 1.9\%(stat.) \\
\hline
  \multirow{2}{*}{6--10}& 0.458 & 0.288 & 0.528 $^{+18\%}_{-14\%}$(scale)  & 0.421 $^{+14\%}_{-10\%}$(scale)  & 0.404 $^{+7.1\%}_{-1.3\%}$(scale) \\
 &  $\pm$ 8.0\%(tot.) & $^{+3.4\%}_{-0.7\%}$(scale) &  \hspace{6mm} $^{+1.9\%}_{-2.4\%}$(pdf) & \hspace{0mm} $\pm$4.7\%(pdf+$\alpha_{s}$)  &  \hspace{4mm}  $\pm$ 4.8\%(stat.) \\
\hline
  \multirow{2}{*}{10--12}& 0.466 & 0.295 & 0.387 $^{+18\%}_{-12\%}$(scale)  & 0.419 $^{+10\%}_{-8.3\%}$(scale)  & 0.353 $^{+3.8\%}_{-2.6\%}$(scale) \\
 &  $\pm$ 8.5\%(tot.) & $^{+10\%}_{-2.4\%}$(scale) &  \hspace{6mm} $^{+2.5\%}_{-1.5\%}$(pdf) & \hspace{0mm} $\pm$4.4\%(pdf+$\alpha_{s}$)  &  \hspace{4mm}  $\pm$ 2.1\%(stat.) \\
\hline
  \multirow{2}{*}{12--14}& 0.451 & 0.310 & 0.352 $^{+16\%}_{-14\%}$(scale)  & 0.419 $^{+9.0\%}_{-7.2\%}$(scale)  & 0.335 $^{+6.3\%}_{-10\%}$(scale) \\
 &  $\pm$ 12\%(tot.) & $^{+8.2\%}_{-5.8\%}$(scale) &  \hspace{6mm} $^{+1.9\%}_{-2.1\%}$(pdf) & \hspace{0mm} $\pm$3.9\%(pdf+$\alpha_{s}$)  &  \hspace{4mm}  $\pm$ 2.6\%(stat.) \\
\hline
  \multirow{2}{*}{14--16}& 0.430 & 0.314 & 0.329 $^{+15\%}_{-13\%}$(scale)  & 0.417 $^{+8.1\%}_{-6.8\%}$(scale)  & 0.338 $^{+6.6\%}_{-9.5\%}$(scale) \\
 &  $\pm$ 15\%(tot.) & $^{+14\%}_{-5.6\%}$(scale) &  \hspace{6mm} $^{+0.9\%}_{-3.6\%}$(pdf) & \hspace{0mm} $\pm$4.2\%(pdf+$\alpha_{s}$)  &  \hspace{4mm}  $\pm$ 2.9\%(stat.) \\
\hline
  \multirow{2}{*}{16--18}& 0.390 & 0.314 & 0.293 $^{+17\%}_{-12\%}$(scale)  & 0.394 $^{+7.4\%}_{-5.5\%}$(scale)  & 0.324 $^{+3.3\%}_{-3.4\%}$(scale) \\
 &  $\pm$ 15\%(tot.) & $^{+13\%}_{-8.7\%}$(scale) &  \hspace{6mm} $^{+1.6\%}_{-2.0\%}$(pdf) & \hspace{0mm} $\pm$4.0\%(pdf+$\alpha_{s}$)  &  \hspace{4mm}  $\pm$ 4.3\%(stat.) \\
\hline
  \multirow{2}{*}{18--20}& 0.354 & 0.298 & 0.254 $^{+18\%}_{-10\%}$(scale)  & 0.354 $^{+7.3\%}_{-5.1\%}$(scale)  & 0.307 $^{+5.7\%}_{-5.0\%}$(scale) \\
 &  $\pm$ 17\%(tot.) & $^{+15\%}_{-11\%}$(scale) &  \hspace{6mm} $^{+3.0\%}_{-0.8\%}$(pdf) & \hspace{0mm} $\pm$4.4\%(pdf+$\alpha_{s}$)  &  \hspace{4mm}  $\pm$ 5.5\%(stat.) \\
\hline
  \multirow{2}{*}{20--22}& 0.336 & 0.277 & 0.226 $^{+16\%}_{-12\%}$(scale)  & 0.320 $^{+5.4\%}_{-6.3\%}$(scale)  & 0.270 $^{+5.4\%}_{-3.4\%}$(scale) \\
 &  $\pm$ 16\%(tot.) & $^{+16\%}_{-10\%}$(scale) &  \hspace{6mm} $^{+1.7\%}_{-2.5\%}$(pdf) & \hspace{0mm} $\pm$3.3\%(pdf+$\alpha_{s}$)  &  \hspace{4mm}  $\pm$ 4.5\%(stat.) \\
\hline
  \multirow{2}{*}{22--24}& 0.289 & 0.255 & 0.201 $^{+14\%}_{-13\%}$(scale)  & 0.278 $^{+9.3\%}_{-5.3\%}$(scale)  & 0.268 $^{+18\%}_{-11\%}$(scale) \\
 &  $\pm$ 14\%(tot.) & $^{+20\%}_{-12\%}$(scale) &  \hspace{6mm} $^{+1.2\%}_{-3.2\%}$(pdf) & \hspace{0mm} $\pm$3.8\%(pdf+$\alpha_{s}$)  &  \hspace{4mm}  $\pm$ 3.3\%(stat.) \\
\hline
  \multirow{2}{*}{24--28}& 0.245 & 0.230 & 0.165 $^{+15\%}_{-10\%}$(scale)  & 0.239 $^{+5.9\%}_{-5.4\%}$(scale)  & 0.223 $^{+8.2\%}_{-9.3\%}$(scale) \\
 &  $\pm$ 14\%(tot.) & $^{+19\%}_{-12\%}$(scale) &  \hspace{6mm} $^{+3.3\%}_{-2.0\%}$(pdf) & \hspace{0mm} $\pm$3.9\%(pdf+$\alpha_{s}$)  &  \hspace{4mm}  $\pm$ 2.7\%(stat.) \\
\hline
  \multirow{2}{*}{28--34}& 0.182 & 0.185 & 0.125 $^{+14\%}_{-12\%}$(scale)  & 0.181 $^{+7.0\%}_{-5.8\%}$(scale)  & 0.178 $^{+7.4\%}_{-6.4\%}$(scale) \\
 &  $\pm$ 15\%(tot.) & $^{+24\%}_{-15\%}$(scale) &  \hspace{6mm} $^{+2.2\%}_{-2.7\%}$(pdf) & \hspace{0mm} $\pm$3.7\%(pdf+$\alpha_{s}$)  &  \hspace{4mm}  $\pm$ 2.8\%(stat.) \\
\hline
  \multirow{2}{*}{34--40}& 0.152 & 0.147 & 0.093 $^{+13\%}_{-11\%}$(scale)  & 0.132 $^{+7.0\%}_{-5.2\%}$(scale)  & 0.136 $^{+3.5\%}_{-2.1\%}$(scale) \\
 &  $\pm$ 12\%(tot.) & $^{+26\%}_{-15\%}$(scale) &  \hspace{6mm} $^{+1.9\%}_{-2.8\%}$(pdf) & \hspace{0mm} $\pm$3.5\%(pdf+$\alpha_{s}$)  &  \hspace{4mm}  $\pm$ 3.2\%(stat.) \\
\hline
  \multirow{2}{*}{40--50}& 0.120 & 0.114 & 0.065 $^{+14\%}_{-11\%}$(scale)  & 0.090 $^{+8.2\%}_{-6.7\%}$(scale)  & 0.107 $^{+14\%}_{-10\%}$(scale) \\
 &  $\pm$ 12\%(tot.) & $^{+28\%}_{-18\%}$(scale) &  \hspace{6mm} $^{+3.1\%}_{-2.7\%}$(pdf) & \hspace{0mm} $\pm$2.8\%(pdf+$\alpha_{s}$)  &  \hspace{4mm}  $\pm$ 1.7\%(stat.) \\
\hline
  \multirow{2}{*}{50--60}& 0.099 & 0.084 & 0.046 $^{+16\%}_{-11\%}$(scale)  & 0.060 $^{+7.8\%}_{-6.7\%}$(scale)  & 0.077 $^{+9.6\%}_{-9.1\%}$(scale) \\
 &  $\pm$ 11\%(tot.) & $^{+34\%}_{-19\%}$(scale) &  \hspace{6mm} $^{+2.5\%}_{-3.9\%}$(pdf) & \hspace{0mm} $\pm$2.4\%(pdf+$\alpha_{s}$)  &  \hspace{4mm}  $\pm$ 2.9\%(stat.) \\
\hline
  \multirow{2}{*}{60--70}& 0.088 & 0.067 & 0.037 $^{+14\%}_{-12\%}$(scale)  & 0.043 $^{+9.8\%}_{-7.5\%}$(scale)  & 0.066 $^{+12\%}_{-13\%}$(scale) \\
 &  $\pm$ 12\%(tot.) & $^{+35\%}_{-21\%}$(scale) &  \hspace{6mm} $^{+3.4\%}_{-4.4\%}$(pdf) & \hspace{0mm} $\pm$3.2\%(pdf+$\alpha_{s}$)  &  \hspace{4mm}  $\pm$ 4.4\%(stat.) \\
\hline
  \multirow{2}{*}{70--80}& 0.070 & 0.054 & 0.030 $^{+16\%}_{-14\%}$(scale)  & 0.034 $^{+9.1\%}_{-7.5\%}$(scale)  & 0.057 $^{+25\%}_{-19\%}$(scale) \\
 &  $\pm$ 14\%(tot.) & $^{+36\%}_{-20\%}$(scale) &  \hspace{6mm} $^{+4.1\%}_{-4.5\%}$(pdf) & \hspace{0mm} $\pm$5.3\%(pdf+$\alpha_{s}$)  &  \hspace{4mm}  $\pm$ 4.0\%(stat.) \\
\hline
  \multirow{2}{*}{80--90}& 0.051 & 0.038 & 0.021 $^{+17\%}_{-14\%}$(scale)  & 0.027 $^{+11\%}_{-9.0\%}$(scale)  & 0.041 $^{+16\%}_{-18\%}$(scale) \\
 &  $\pm$ 12\%(tot.) & $^{+38\%}_{-22\%}$(scale) &  \hspace{6mm} $^{+4.7\%}_{-4.9\%}$(pdf) & \hspace{0mm} $\pm$3.5\%(pdf+$\alpha_{s}$)  &  \hspace{4mm}  $\pm$ 5.8\%(stat.) \\
\hline
  \multirow{2}{*}{90--100}& 0.033 & 0.025 & 0.014 $^{+22\%}_{-13\%}$(scale)  & 0.016 $^{+12\%}_{-6.9\%}$(scale)  & 0.030 $^{+2.3\%}_{-4.0\%}$(scale) \\
 &  $\pm$ 14\%(tot.) & $^{+44\%}_{-21\%}$(scale) &  \hspace{6mm} $^{+9.1\%}_{-3.6\%}$(pdf) & \hspace{0mm} $\pm$4.4\%(pdf+$\alpha_{s}$)  &  \hspace{4mm}  $\pm$ 7.0\%(stat.) \\
\hline
  \multirow{2}{*}{100--120}& 0.018 & 0.015 & 0.008 $^{+21\%}_{-14\%}$(scale)  & 0.010 $^{+9.2\%}_{-10\%}$(scale)  & 0.017 $^{+8.4\%}_{-9.1\%}$(scale) \\
 &  $\pm$ 12\%(tot.) & $^{+44\%}_{-24\%}$(scale) &  \hspace{6mm} $^{+9.3\%}_{-6.2\%}$(pdf) & \hspace{0mm} $\pm$6.2\%(pdf+$\alpha_{s}$)  &  \hspace{4mm}  $\pm$ 7.6\%(stat.) \\
\hline
  \multirow{2}{*}{120--200}& 0.0044 & 0.0039 & 0.0023 $^{+24\%}_{-17\%}$(scale)  & 0.0032 $^{+12\%}_{-9.0\%}$(scale)  & 0.0043 $^{+18\%}_{-8.0\%}$(scale) \\
 &  $\pm$ 13\%(tot.) & $^{+48\%}_{-26\%}$(scale) &  \hspace{6mm} $^{+14\%}_{-13\%}$(pdf) & \hspace{0mm} $\pm$7.1\%(pdf+$\alpha_{s}$)  &  \hspace{4mm}  $\pm$ 14\%(stat.) \\
\hline
\end{tabular}
}
\label{tab:2gNNLOPredictionQt}
\end{table*}

\begin{table*}[htbp]
\centering
\topcaption{Values of $\rd\sigma/\rd\Delta \phi_{\Pgg\Pgg}$ (pb/rad) for the data, \SHERPA, \DIPHOX+\GENERATORGTOMC, \RESBOS and \TWOGNNLO predictions.}
{\footnotesize
\begin{tabular}{c|ccccc}
\hline
        &  \multicolumn{5}{c}{ $d\sigma/d\Delta \phi_{\Pgg\Pgg}$ (pb/rad)} \\
\hline
\dphigg & Data  &  \SHERPA & \DIPHOX & \RESBOS  &  \TWOGNNLO \\
(rad)&   &   & +\GENERATORGTOMC & & \\
\hline
  \multirow{2}{*}{(0.00--0.20)$\pi$ }& 0.92 & 0.58 & 0.36 $^{+24\%}_{-16\%}$(scale)  & 0.41 $^{+9.9\%}_{-7.6\%}$(scale)  & 0.71 $^{+17\%}_{-18\%}$(scale) \\
 &  $\pm$ 19\%(tot.) & $^{+42\%}_{-22\%}$(scale) &  \hspace{6mm} $^{+7.2\%}_{-6.9\%}$(pdf) & \hspace{0mm} $\pm$15\%(pdf+$\alpha_{s}$)  &  \hspace{4mm}  $\pm$ 3.5\%(stat.) \\
\hline
  \multirow{2}{*}{(0.20--0.40)$\pi$ }& 1.42 & 0.85 & 0.45 $^{+21\%}_{-14\%}$(scale)  & 0.44 $^{+11\%}_{-9\%}$(scale)  & 0.98 $^{+12\%}_{-18\%}$(scale) \\
 &  $\pm$ 12\%(tot.) & $^{+43\%}_{-24\%}$(scale) &  \hspace{6mm} $^{+7.0\%}_{-5.3\%}$(pdf) & \hspace{0mm} $\pm$3.1\%(pdf+$\alpha_{s}$)  &  \hspace{4mm}  $\pm$ 3.0\%(stat.) \\
\hline
  \multirow{2}{*}{(0.40--0.60)$\pi$ }& 2.06 & 1.50 & 0.68 $^{+15\%}_{-14\%}$(scale)  & 0.76 $^{+9.8\%}_{-7.4\%}$(scale)  & 1.38 $^{+12\%}_{-12\%}$(scale) \\
 &  $\pm$ 11\%(tot.) & $^{+39\%}_{-20\%}$(scale) &  \hspace{6mm} $^{+3.6\%}_{-6.0\%}$(pdf) & \hspace{0mm} $\pm$2.8\%(pdf+$\alpha_{s}$)  &  \hspace{4mm}  $\pm$ 2.3\%(stat.) \\
\hline
  \multirow{2}{*}{(0.60--0.70)$\pi$ }& 3.42 & 2.69 & 1.21 $^{+17\%}_{-11\%}$(scale)  & 1.60 $^{+8.5\%}_{-6.5\%}$(scale)  & 2.33 $^{+10\%}_{-10\%}$(scale) \\
 &  $\pm$ 10\%(tot.) & $^{+31\%}_{-19\%}$(scale) &  \hspace{6mm} $^{+4.1\%}_{-3.3\%}$(pdf) & \hspace{0mm} $\pm$3.0\%(pdf+$\alpha_{s}$)  &  \hspace{4mm}  $\pm$ 2.6\%(stat.) \\
\hline
  \multirow{2}{*}{(0.70--0.80)$\pi$ }& 5.64 & 4.43 & 2.32 $^{+15\%}_{-11\%}$(scale)  & 3.21 $^{+7.4\%}_{-6.1\%}$(scale)  & 4.02 $^{+8.1\%}_{-8.1\%}$(scale) \\
 &  $\pm$ 9.8\%(tot.) & $^{+29\%}_{-17\%}$(scale) &  \hspace{6mm} $^{+2.5\%}_{-2.9\%}$(pdf) & \hspace{0mm} $\pm$3.0\%(pdf+$\alpha_{s}$)  &  \hspace{4mm}  $\pm$ 1.8\%(stat.) \\
\hline
  \multirow{2}{*}{(0.80--0.84)$\pi$ }& 8.95 & 6.85 & 4.06 $^{+15\%}_{-12\%}$(scale)  & 5.75 $^{+7.4\%}_{-6.1\%}$(scale)  & 7.01 $^{+19\%}_{-10\%}$(scale) \\
 &  $\pm$ 10\%(tot.) & $^{+24\%}_{-15\%}$(scale) &  \hspace{6mm} $^{+2.1\%}_{-2.5\%}$(pdf) & \hspace{0mm} $\pm$3.3\%(pdf+$\alpha_{s}$)  &  \hspace{4mm}  $\pm$ 3.2\%(stat.) \\
\hline
  \multirow{2}{*}{(0.84--0.88)$\pi$ }& 10.9 & 9.3 & 6.3 $^{+15\%}_{-11\%}$(scale)  & 8.56 $^{+7.8\%}_{-5.9\%}$(scale)  & 9.41 $^{+9.0\%}_{-11\%}$(scale) \\
 &  $\pm$ 11\%(tot.) & $^{+20\%}_{-15\%}$(scale) &  \hspace{6mm} $^{+2.1\%}_{-2.3\%}$(pdf) & \hspace{0mm} $\pm$3.5\%(pdf+$\alpha_{s}$)  &  \hspace{4mm}  $\pm$ 3.7\%(stat.) \\
\hline
  \multirow{2}{*}{(0.88--0.90)$\pi$ }& 14.4 & 12.0 & 9.1 $^{+15\%}_{-12\%}$(scale)  & 12.0 $^{+7.6\%}_{-5.9\%}$(scale)  & 12.2 $^{+7.7\%}_{-11\%}$(scale) \\
 &  $\pm$ 11\%(tot.) & $^{+19\%}_{-13\%}$(scale) &  \hspace{6mm} $^{+4.3\%}_{-1.7\%}$(pdf) & \hspace{0mm} $\pm$3.6\%(pdf+$\alpha_{s}$)  &  \hspace{4mm}  $\pm$ 4.2\%(stat.) \\
\hline
  \multirow{2}{*}{(0.90--0.92)$\pi$ }& 16.9 & 14.0 & 12.3 $^{+14\%}_{-12\%}$(scale)  & 15.3 $^{+8.2\%}_{-6.6\%}$(scale)  & 14.1 $^{+8.3\%}_{-3.8\%}$(scale) \\
 &  $\pm$ 11\%(tot.) & $^{+18\%}_{-11\%}$(scale) &  \hspace{6mm} $^{+2.8\%}_{-2.3\%}$(pdf) & \hspace{0mm} $\pm$3.5\%(pdf+$\alpha_{s}$)  &  \hspace{4mm}  $\pm$ 11\%(stat.) \\
\hline
  \multirow{2}{*}{(0.92--0.94)$\pi$ }& 21.4 & 17.1 & 17.0 $^{+19\%}_{-12\%}$(scale)  & 20.8 $^{+8.6\%}_{-6.9\%}$(scale)  & 18.7 $^{+7.4\%}_{-2.4\%}$(scale) \\
 &  $\pm$ 11\%(tot.) & $^{+16\%}_{-9.0\%}$(scale) &  \hspace{6mm} $^{+3.6\%}_{-1.8\%}$(pdf) & \hspace{0mm} $\pm$3.9\%(pdf+$\alpha_{s}$)  &  \hspace{4mm}  $\pm$ 5.9\%(stat.) \\
\hline
  \multirow{2}{*}{(0.94--0.96)$\pi$ }& 24.9 & 20.7 & 25.0 $^{+16\%}_{-12\%}$(scale)  & 28.0 $^{+9.8\%}_{-7.6\%}$(scale)  & 23.4 $^{+8.2\%}_{-2.3\%}$(scale) \\
 &  $\pm$ 11\%(tot.) & $^{+14\%}_{-6.1\%}$(scale) &  \hspace{6mm} $^{+6.2\%}_{-1.7\%}$(pdf) & \hspace{0mm} $\pm$3.8\%(pdf+$\alpha_{s}$)  &  \hspace{4mm}  $\pm$ 3.1\%(stat.) \\
\hline
  \multirow{2}{*}{(0.96--0.98)$\pi$ }& 29.1 & 26.2 & 43.3 $^{+17\%}_{-14\%}$(scale)  & 38.6 $^{+13\%}_{-10\%}$(scale)  & 34.3 $^{+3.2\%}_{-6.9\%}$(scale) \\
 &  $\pm$ 12\%(tot.) & $^{+10\%}_{-6.2\%}$(scale) &  \hspace{6mm} $^{+1.5\%}_{-2.1\%}$(pdf) & \hspace{0mm} $\pm$4.0\%(pdf+$\alpha_{s}$)  &  \hspace{4mm}  $\pm$ 3.1\%(stat.) \\
\hline
  \multirow{2}{*}{(0.98--1.00)$\pi$ }& 38.2 & 32.2 & 44.3 $^{+7.4\%}_{-14\%}$(scale)  & 54.1 $^{+32\%}_{-23\%}$(scale)  & 54.6 $^{+8.0\%}_{-7.3\%}$(scale) \\
 &  $\pm$ 11\%(tot.) & $^{+7.1\%}_{-2.3\%}$(scale) &  \hspace{6mm} $^{+0.8\%}_{-4.6\%}$(pdf) & \hspace{0mm} $\pm$4.3\%(pdf+$\alpha_{s}$)  &  \hspace{4mm}  $\pm$ 2.0\%(stat.) \\
\hline
\end{tabular}
}
\label{tab:2gNNLOPredictionDphi}
\end{table*}

\begin{table*}[htbp]
\centering
\topcaption{Values of $\rd\sigma/\rd\abs{\cos\theta^{*}}$ (pb) for the data, \SHERPA, \DIPHOX+\GENERATORGTOMC, \RESBOS and \TWOGNNLO predictions.}
{\footnotesize

\begin{tabular}{c|ccccc}
\hline
      &  \multicolumn{5}{c}{ $\rd\sigma/\rd\abs{\cos\theta^{*}}$ (pb)} \\
\hline
$\abs{\cos\theta^{*}}$& Data  &  \SHERPA & \DIPHOX & \RESBOS  &  \TWOGNNLO \\
&   &   & +\GENERATORGTOMC & & \\
\hline
  \multirow{2}{*}{0.00--0.20}& 22.3 & 16.0 & 16.0 $^{+11\%}_{-11\%}$(scale)  & 21.1 $^{+14\%}_{-11\%}$(scale)  & 21.5 $^{+6.5\%}_{-5.4\%}$(scale) \\
 &  $\pm$ 11\%(tot.) & $^{+18\%}_{-11\%}$(scale) &  \hspace{6mm} $^{+2.6\%}_{-6.5\%}$(pdf) & \hspace{0mm} $\pm$3.7\%(pdf+$\alpha_{s}$)  &  \hspace{4mm}  $\pm$ 2.0\%(stat.) \\
\hline
  \multirow{2}{*}{0.20--0.28}& 19.8 & 16.1 & 15.3 $^{+11\%}_{-10\%}$(scale)  & 19.6 $^{+14\%}_{-11\%}$(scale)  & 20.3 $^{+3.7\%}_{-6.4\%}$(scale) \\
 &  $\pm$ 11\%(tot.) & $^{+18\%}_{-10\%}$(scale) &  \hspace{6mm} $^{+5.0\%}_{-4.4\%}$(pdf) & \hspace{0mm} $\pm$3.8\%(pdf+$\alpha_{s}$)  &  \hspace{4mm}  $\pm$ 2.9\%(stat.) \\
\hline
  \multirow{2}{*}{0.28--0.36}& 20.0 & 15.8 & 15.0 $^{+13\%}_{-10\%}$(scale)  & 18.6 $^{+14\%}_{-11\%}$(scale)  & 18.8 $^{+5.5\%}_{-5.7\%}$(scale) \\
 &  $\pm$ 10\%(tot.) & $^{+19\%}_{-10\%}$(scale) &  \hspace{6mm} $^{+4.4\%}_{-5.3\%}$(pdf) & \hspace{0mm} $\pm$3.8\%(pdf+$\alpha_{s}$)  &  \hspace{4mm}  $\pm$ 2.9\%(stat.) \\
\hline
  \multirow{2}{*}{0.36--0.44}& 18.8 & 15.6 & 14.6 $^{+11\%}_{-10\%}$(scale)  & 17.6 $^{+14\%}_{-11\%}$(scale)  & 18.0 $^{+9.0\%}_{-9.5\%}$(scale) \\
 &  $\pm$ 9.7\%(tot.) & $^{+19\%}_{-12\%}$(scale) &  \hspace{6mm} $^{+4.8\%}_{-4.3\%}$(pdf) & \hspace{0mm} $\pm$3.8\%(pdf+$\alpha_{s}$)  &  \hspace{4mm}  $\pm$ 3.1\%(stat.) \\
\hline
  \multirow{2}{*}{0.44--0.60}& 18.4 & 14.9 & 13.7 $^{+11\%}_{-9.3\%}$(scale)  & 16.3 $^{+15\%}_{-11\%}$(scale)  & 17.9 $^{+9.9\%}_{-6.4\%}$(scale) \\
 &  $\pm$ 9.9\%(tot.) & $^{+21\%}_{-11\%}$(scale) &  \hspace{6mm} $^{+5.4\%}_{-4.5\%}$(pdf) & \hspace{0mm} $\pm$3.7\%(pdf+$\alpha_{s}$)  &  \hspace{4mm}  $\pm$ 2.1\%(stat.) \\
\hline
  \multirow{2}{*}{0.60--0.90}& 13.7 & 12.2 & 10.9 $^{+11\%}_{-9.2\%}$(scale)  & 10.9 $^{+15\%}_{-11\%}$(scale)  & 12.9 $^{+3.9\%}_{-6.8\%}$(scale) \\
 &  $\pm$ 11\%(tot.) & $^{+22\%}_{-12\%}$(scale) &  \hspace{6mm} $^{+4.3\%}_{-5.4\%}$(pdf) & \hspace{0mm} $\pm$3.7\%(pdf+$\alpha_{s}$)  &  \hspace{4mm}  $\pm$ 2.7\%(stat.) \\
\hline
  \multirow{2}{*}{0.90--1.00}& 10.4 & 6.7 & 5.6 $^{+8.0\%}_{-13\%}$(scale)  & 3.6 $^{+14\%}_{-11\%}$(scale)  & 6.5 $^{+13\%}_{-17\%}$(scale) \\
 &  $\pm$ 21\%(tot.) & $^{+32\%}_{-16\%}$(scale) &  \hspace{6mm} $^{+2.9\%}_{-6.4\%}$(pdf) & \hspace{0mm} $\pm$6.2\%(pdf+$\alpha_{s}$)  &  \hspace{4mm}  $\pm$ 5.5\%(stat.) \\
\hline
\end{tabular}
}
\label{tab:2gNNLOPredictionCostt}
\end{table*}

\cleardoublepage \section{The CMS Collaboration \label{app:collab}}\begin{sloppypar}\hyphenpenalty=5000\widowpenalty=500\clubpenalty=5000\textbf{Yerevan Physics Institute,  Yerevan,  Armenia}\\*[0pt]
S.~Chatrchyan, V.~Khachatryan, A.M.~Sirunyan, A.~Tumasyan
\vskip\cmsinstskip
\textbf{Institut f\"{u}r Hochenergiephysik der OeAW,  Wien,  Austria}\\*[0pt]
W.~Adam, T.~Bergauer, M.~Dragicevic, J.~Er\"{o}, C.~Fabjan\cmsAuthorMark{1}, M.~Friedl, R.~Fr\"{u}hwirth\cmsAuthorMark{1}, V.M.~Ghete, C.~Hartl, N.~H\"{o}rmann, J.~Hrubec, M.~Jeitler\cmsAuthorMark{1}, W.~Kiesenhofer, V.~Kn\"{u}nz, M.~Krammer\cmsAuthorMark{1}, I.~Kr\"{a}tschmer, D.~Liko, I.~Mikulec, D.~Rabady\cmsAuthorMark{2}, B.~Rahbaran, H.~Rohringer, R.~Sch\"{o}fbeck, J.~Strauss, A.~Taurok, W.~Treberer-Treberspurg, W.~Waltenberger, C.-E.~Wulz\cmsAuthorMark{1}
% ...
\vskip\cmsinstskip
% ...
1:~~Also at Vienna University of Technology, Vienna, Austria\\
2:~~Also at CERN, European Organization for Nuclear Research, Geneva, Switzerland\\
% ...

\end{sloppypar}
\end{document}